\documentclass[twocolumn,a4paper,amsmath,amssymb,showpacs,prb,superscriptaddress]{revtex4-1}
 
\usepackage[dvipdfmx]{graphicx}
\usepackage{bmpsize}
\usepackage{amssymb, amsfonts, amsmath}
\usepackage{dcolumn}
\usepackage{bm}
\usepackage{color}
\usepackage[normalem]{ulem}

\begin{document}
\title{Multipole expansion for magnetic structures: A generation scheme for symmetry-adapted orthonormal basis set in crystallographic point group}
\author{M.-T. Suzuki}
\affiliation{Center for Computational Materials Science, Institute for Materials Research, Tohoku University, Sendai, Miyagi 980-8577, Japan}
\author{T. Nomoto}
\affiliation{Department of Applied Physics, University of Tokyo 7-3-1 Hongo Bunkyo-ku, Tokyo 113-8656, Japan}
\author{R. Arita}
\affiliation{Department of Applied Physics, University of Tokyo 7-3-1 Hongo Bunkyo-ku, Tokyo 113-8656, Japan}
\affiliation{RIKEN Center for Emergent Matter Science (CEMS), Wako, Saitama 351-0198, Japan}
\author{Y. Yanagi}
\affiliation{Center for Computational Materials Science, Institute for Materials Research, Tohoku University, Sendai, Miyagi 980-8577, Japan}
\author{S. Hayami}
\affiliation{Department of Physics, Hokkaido University, Sapporo 060-0810, Japan}
\author{H. Kusunose}
\affiliation{Department of Physics, Meiji University, Kawasaki 214-8571, Japan}
\date{\today}

\begin{abstract}
We propose a systematic method to generate a complete orthonormal basis set of multipole expansion for magnetic structures in arbitrary crystal structure.
The key idea is the introduction of a virtual atomic cluster of a target crystal, on which we can clearly define the magnetic configurations corresponding to symmetry-adapted multipole moments.
The magnetic configurations are then mapped onto the crystal so as to preserve the magnetic point group of the multipole moments, leading to the magnetic structures classified according to the irreducible representations of crystallographic point group.
We apply the present scheme to pyrhochlore and hexagonal $AB$O$_3$ crystal structures, and demonstrate that the multipole expansion is useful to investigate the macroscopic responses of antiferromagnets.
\end{abstract}

\maketitle

\section{INTRODUCTION}
Diversity of physical properties of magnets provides a fascinating playground in condensed matter physics.
When we explore this exciting arena, what is interesting to note is that there are many restrictions imposed by the symmetry of the magnetic structures.
For example, it has been well known that the structure of linear response tensors are determined by the magnetic point group\cite{Birss1962,Birss1964,Kleiner1966,Rivera2009,Szaller2013,Seemann2015,Suzuki2017}.
Furthermore, identifying the order parameter for the magnetic phase is most useful for deeper understanding of physical phenomena.
There has been many studies to investigate the relation between the
order parameters in a particular magnetic structure and macroscopic
phenomena such as anomalous Hall (AH)
effect~\cite{Shindou2001,Suzuki2017,Smejkal2019}, electromagnetic (EM)
effect~\cite{Ederer2007,Spaldin2008,Spaldin2013,Hayami2014,Hayami2018,Watanabe2018}, and optical responses~\cite{Higo2018}.
Thus we have a significant chance to specify or even design a magnet exhibiting desired physical properties by investigating the order parameters which characterize the magnetic structures.

For this purpose, the multipole expansion of the magnetic structure is an efficient and powerful approach.
Indeed, the multipole moments inherent in the magnetic structure, which we call cluster multipole moments~\cite{Suzuki2017}, have played a crucial role as a key order parameter in a variety of studies: EM effect has been discussed in terms of the magnetic (M) rank-0 monopole and rank-2 quadrupole as well as the magnetic toroidal (MT) rank-1 dipole as the order parameter to characterize the specific magnetic structures~\cite{Ederer2007,Spaldin2008,Spaldin2013,Thoele2016}.
The relation between parity odd multipoles and electromagnetic effect
has recently been investigated based on generalized forms of the multipole expansions for magnetic distributions~\cite{Hayami2018,Watanabe2018}.
It has also been shown that the rank-3 M multipole (octupole) plays a key role for a large AH effect~\cite{Nakatsuji2015,Kiyohara2016,Nayak2016}, anomalous Nernst effect~\cite{Ikhlas2017}, and magneto-optical Kerr effect~\cite{Higo2018} in the coplanar antiferromagnets Mn$_3$$Z$ ($Z$=Sn, Ge)~\cite{Suzuki2017}.
In these studies, the interplay between the physical properties and the magnetic structure through cluster multipoles has been investigated extensively.
However, there has been no concrete scheme to make a complete basis set of cluster multipoles for a given crystal structure.

In this paper, we propose a scheme to generate cluster multipoles which form a complete orthonormal basis set for arbitrary magnetic structures.
Here, we introduce a virtual atomic cluster, which depends only on the crystallographic point group of the system.
We define unambiguously the magnetic configurations corresponding to the symmetry-adapted multipoles in the atomic cluster.
The obtained magnetic configurations are mapped to the original crystal structure with the magnetic point group symmetry preserved.
The generated complete basis set for the magnetic structure in crystal is useful to measure the symmetry breaking as an order parameter according to the magnetic point group~\cite{Suzuki2017}.
Although we restrict ourselves to the case with ``uniform'' magnetic structures characterized by the ordering vector ${\bm q}=0$ in this paper, an extension to cases with nonzero $\bm{q}$ ordering vectors is straightforward.

To demonstrate the efficiency of the present scheme, we apply the cluster multipole expansion of magnetic structures to pyrochlore and hexagonal $AB$O$_3$ crystal structures.
For the pyrochlore structure, it is shown that the all-in all-out magnetic structure corresponds to a 
M octupole, and two-in two-out and one-in three-out magnetic structures are expressed by the linear combinations of the M dipole and octupole that 
belong to the same irreducible representation (IREP) of the crystallographic point group.
The two-in two-out and one-in three-out magnetic structures can be transformed continuously to the pure antiferromagnetic structures, indicating that the antiferromagnetic structures without net magnetization yield the AH effect.
For the hexagonal $AB$O$_3$ structure, higher rank multipoles such as MT quadrupole and M octupole are necessary to describe the magnetic structures exhibiting the AH and EM effects within the uniform magnetic ordering.

\begin{center}
\begin{figure*}[t]
\includegraphics[width=0.9\linewidth]{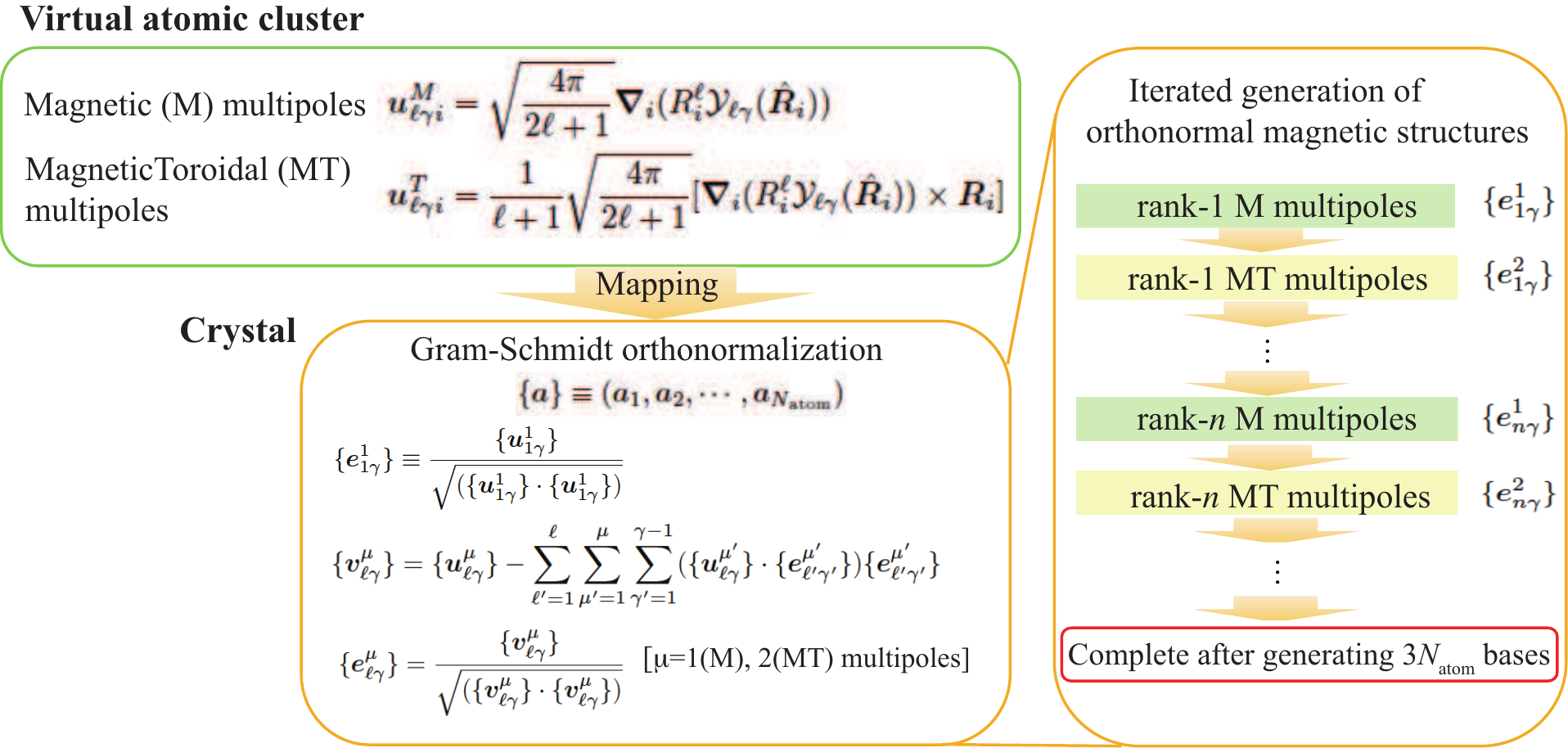}
\caption{Outline of the generation procedure for 3$N_{{\rm atom}}$ complete basis set of multipole magnetic structure classified by the crystallographic point group.}
\label{Fig:GenerationFlow}
\end{figure*}
\end{center}

\section{Multipole expansion of magnetic structures in crystal}
\subsection{Outline of multipole expansion of magnetic structures in crystal}
In this paper, we present a method to generate an orthonormal complete basis set of magnetic structures corresponding to the cluster multipoles classified according to the IREPs of the crystallographic point group.
To make free from confusion and ambiguity, we define some words used in this paper.
``Atomic cluster'' is used for the atoms transformed to each other only by the rotation operations of a point group and distinguished from ``crystal'' which assumes a periodicity for an atomic configuration.
``Magnetic configuration'' is used for the alignment of magnetic dipole moments on atoms of the atomic cluster and distinguished from ``magnetic structure'' which is for the alignment of magnetic dipole moments on atoms in periodic crystal.
A magnetic configuration (a magnetic structure) characterized by the M and/or MT multipoles is called multipole magnetic configuration (multipole magnetic structure).
The outline of the generation procedure for the orthonormal multipole
magnetic structures in crystal is following (see also Fig.~\ref{Fig:GenerationFlow}):

\begin{enumerate}
\item Set a virtual atomic cluster corresponding to the crystallographic point group of a target crystal.
\item Generate magnetic configurations corresponding to the symmetrized M and MT multipoles in the virtual atomic cluster.
\item Map the magnetic dipole moments on the atoms in the virtual atomic cluster to those on the crystallographically equivalent atoms in crystal, which represent the multipole magnetic structures.
\item Orthonormalize the bases of multipole magnetic structures by using
      the Gram-Schmidt orthogonalization procedure.
\end{enumerate}

Details for the procedure will be explained in Sec.~\ref{Sec:VirtualCluster}, \ref{Sec:Mapping}, and \ref{Sec:Orthonormalization}.
With this method, all the generated orthonormal magnetic structures are characterized by the symmetry-adapted multipoles.
Thus, the multipoles can be a useful measure of symmetry breaking of the crystallographic point group in the presence of the uniform magnetic orders.
It is a natural extension of the conventional dipole magnetization which measures the symmetry breaking of the ferromagnetic order.
The multipole expansion for magnetic structures in crystal is very efficient to investigate the relation between the magnetic structures and physical properties beyond the symmetry analysis.

\subsection{Multipole magnetic configurations in atomic clusters}
The multipole expansion of the vector gauge potential is given under the Coulomb gauge $\bm{\nabla}\cdot{\bm A}({\bm r})=0$ as follows:
\begin{align}
{\bm A}({\bm r})=\sum_{\ell m}\biggl(b_{\ell}M_{\ell m} \frac{{\bm Y}_{\ell m}^{\ell}(\hat{{\bm r}})}{r^{\ell+1}} + c_{\ell} T_{\ell m}\frac{{\bm Y}_{\ell m}^{\ell+1}(\hat{{\bm r}})}{r^{\ell+2}} \biggr),
\label{Eq:VecPot}
\end{align}
where ${\bm Y}_{\ell m}^{\ell'}(\hat{{\bm r}})$ ($\ell\ge 1$, $-\ell\le m \le \ell$, $\ell'=\ell-1, \ell,\ell+1$) is the vector spherical harmonics that transforms as conventional scalar spherical harmonics $Y_{\ell m}(\hat{{\bm r}})$ for rotation operation with its orbital angular momentum $\ell'$~\cite{Blatt1991,Kusunose2008,Varshalovich1988}.
The expansion coefficients in Eq.~(\ref{Eq:VecPot}) 
are the so-called M multipole, $M_{\ell m}$, and MT multipole, $T_{\ell m}$, respectively ($b_{\ell}$ and $c_{\ell}$ are introduced for convenience).
The M and MT multipoles around a single magnetic ion in the unit of Bohr magneton are defined as
\begin{align}
M_{\ell m}&=\sqrt{\frac{4\pi}{2\ell+1}}\sum_{j}\biggl(\frac{2{\bm\ell}_j}{\ell +1}+{\bm \sigma_j}\biggr)\cdot \bm{O}_{lm}(\bm{r}_{j}),
\label{Eq:FullMP}
\\
T_{\ell m}&=\sqrt{\frac{4\pi}{2\ell+1}}\sum_{j}\biggl\{\frac{{\bm r}_j}{\ell +1}\times\biggl(\frac{2{\bm \ell}_j}{\ell +2}+{\bm \sigma}_j\biggr)\biggr\} \cdot
\bm{O}_{lm}(\bm{r}_{j}),
\label{Eq:FullTD}
\end{align}
with
\begin{align}
\bm{O}_{lm}(\bm{r})\equiv \bm{\nabla}\left[ r^{\ell}Y^{*}_{\ell m}(\hat{\bm{r}}) \right],
\label{eq:opry}
\end{align}
where ${\bm \ell}_j$ and ${\bm\sigma}_j$ are the orbital and spin angular momentum of an electron at ${\bm r}_j$.
Here, $r=|\bm{r}|$ and $\hat{\bm{r}}=\bm{r}/r$.
Since the vector spherical harmonics ${\bm Y}_{\ell m}^{\ell'}$ are an orthonormal complete basis set of a vector function on a sphere, the M and MT multipoles $M_{\ell m}$ and $T_{\ell m}$ represent the arbitrary angular dependence of the magnetization distribution on a single magnetic ion.
Focusing on the spin part of Eqs.~(\ref{Eq:FullMP}) and (\ref{Eq:FullTD}), the M and MT multipoles can be extended straightforwardly to characterize the classical magnetic configurations $\{{\bm m}_i \}$ in an atomic cluster whose atoms are transformed to each other through the point-group symmetry operations with respect to the symmetry center of the atomic cluster.
The explicit definitions are given as
\begin{align}
M_{\ell m}&\equiv \sqrt{\frac{4\pi}{2\ell+1}}\sum_{i=1}^{N_{\rm atom}}{\bm m}_i\cdot
\bm{O}_{lm}(\bm{R}_{i}), \\
T_{\ell m}&\equiv \frac{1}{\ell+1}\sqrt{\frac{4\pi}{2\ell+1}}\sum_{i=1}^{N_{\rm atom}}({\bm R}_i\times {\bm m}_{i})\cdot
\bm{O}_{lm}(\bm{R}_{i}),
\end{align}
where ${\bm R}_{i}$ is the position vector of $i$-th atom and $N_{\rm atom}$ is the number of atoms in the atomic cluster.
These multipoles have the same transformation property with that of the spherical harmonics $Y_{\ell m}$ for rotation operation of the point group, and hence they are classified according to the IREPs of the point group~\cite{Hayami2018}.
The spherical harmonics are usually symmetrized according to the IREPs as
\begin{align}
\mathcal{Y}_{\ell \gamma}(\hat{{\bm r}}) =\sum_{m} c^{\gamma}_{\ell m} Y_{\ell m}(\hat{{\bm r}}),
\label{Eq:Spherical}
\end{align}
where $\gamma$ runs from 1 to $2\ell+1$ in order to distinguish the IREP and its component including multiplicity which is necessary when the same IREPs are multiply appeared in the same rank $\ell$.
The coefficients $c^{\gamma}_{\ell m}$ of the symmetrized spherical harmonics $\mathcal{Y}_{\ell \gamma}(\hat{{\bm r}})$ are tabulated in Ref.~\onlinecite{Kusunose2008} for instance, where $c^{\gamma}_{\ell m}$ are chosen so that $\mathcal{Y}_{\ell \gamma}(\hat{{\bm r}})$ is real.
This is always possible in the presence of the time-reversal symmetry.
The symmetry-adapted multipoles are thus reexpressed as follows:
\begin{align}
M_{\ell \gamma}&=\sum_{i=1}^{N_{\rm atom}} {\bm u}_{\ell\gamma i}^{M}\cdot {\bm m}_i,
\label{Eq:CMP} \\
T_{\ell \gamma}&=\sum_{i=1}^{N_{\rm atom}} {\bm u}_{\ell\gamma i}^{T}\cdot {\bm m}_i,
\label{Eq:CTD}
\end{align}
where
\begin{align}
{\bm u}^{M}_{\ell\gamma i}&=\sqrt{\frac{4\pi}{2\ell+1}}\bm{\mathcal{O}}_{l\gamma}(\bm{R}_i),
\label{Eq:base_mp} \\
{\bm u}^{T}_{\ell\gamma i}&=\frac{1}{\ell+1}\sqrt{\frac{4\pi}{2\ell+1}}
\left(\bm{\mathcal{O}}_{l\gamma}(\bm{R}_i)\times {\bm R}_i\right),
\label{Eq:base_td}
\end{align}
and $\bm{\mathcal{O}}_{l\gamma}$ is given by replacing $Y_{lm}^{*}$ in Eq.~(\ref{eq:opry}) with $\mathcal{Y}_{l\gamma}$.

In generating an orthogonal basis set of the magnetic structures based on the symmetry, the conventional projection operator method requires nontrivial trials to find out the suitable trial functions, and become complicated when the crystal contains many magnetic atoms.
Moreover, that procedure has large ambiguity for low-symmetry crystals, and for the complicated magnetic structures, it would be difficult to find out the correspondence between the magnetic structures and symmetry-adapted multipoles.
On the contrary, our new method is highly advantageous since it is possible to automatically generate a complete orthonormal basis set of magnetic structures, and it is, by definition, based on the symmetry-adapted multipoles.

Here, we comment on the so-called M monopole or magnetic flux configuration that corresponds to a magnetic configuration in which all the magnetic dipole moments point to the center of the atomic cluster.
 The monopole magnetic configuration is parity odd and invariant
under all the rotational symmetry operations of the point group. 
Such a monopole magnetic configuration can be defined as~\cite{Spaldin2013,Thole2016}:
\begin{align}
{\bm u}^{M}_{01i}\equiv\frac{{\bm R}_i}{R_i^{2}}.
\label{Eq:base_mono}
\end{align}
In this paper, however, we do not include this type of M monopole in the cluster multipole expansion since the ordinary multipole expansion in Eq.~(\ref{Eq:VecPot}) does not contain the monopole term and the corresponding magnetic configuration always appears as a higher-rank multipole of the parity odd fully symmetric 
IREP, as will be shown in the case of hexagonal $AB$O$_3$ in Sec.~\ref{Sec:HexABO3}.

\subsection{Virtual atomic clusters for crystallographic point groups}
\label{Sec:VirtualCluster}

\begin{figure}[tb]
\includegraphics[width=0.8\linewidth]{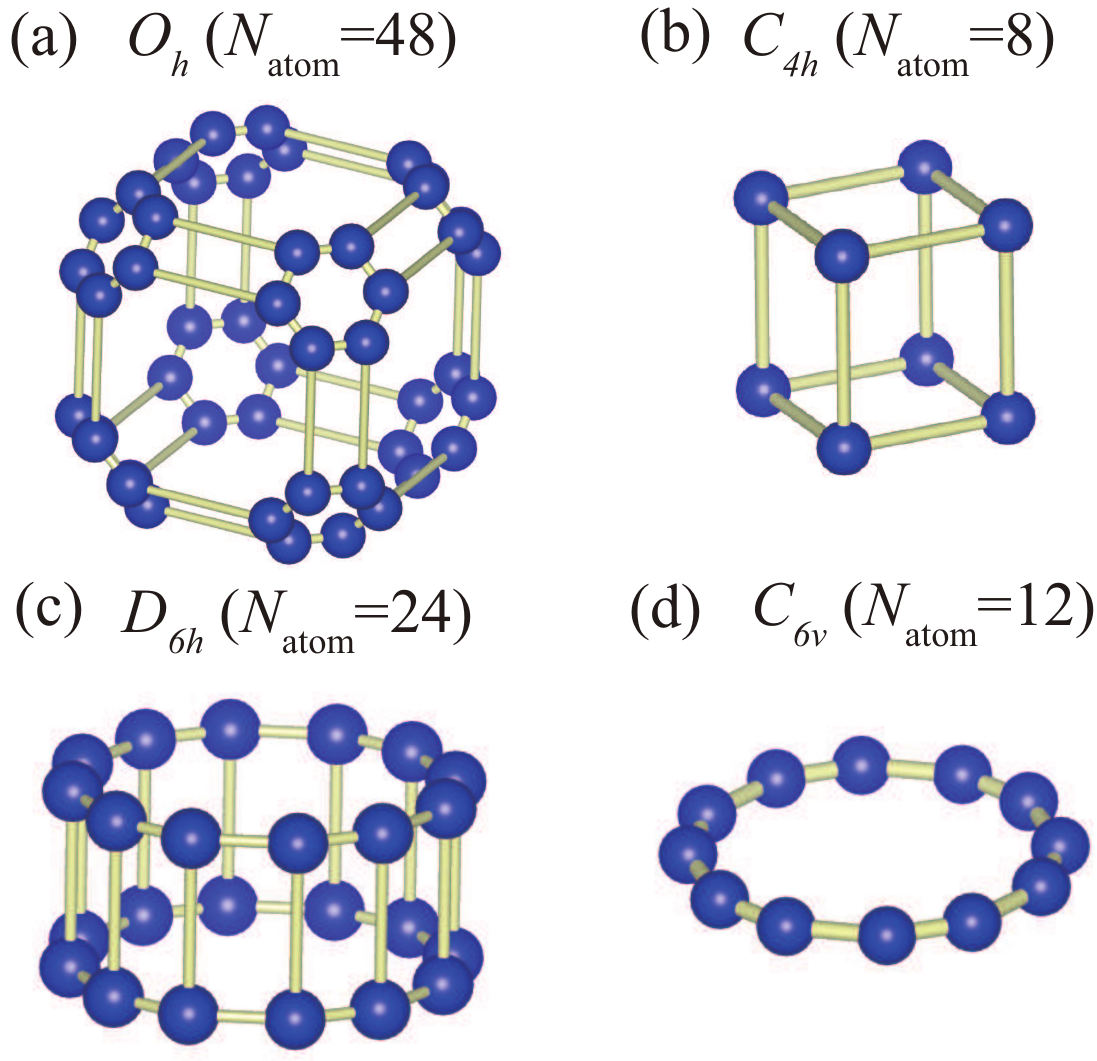}
\caption{Examples of virtual atomic clusters for (a) $O_{h}$, (b) $C_{4h}$, (c) $D_{6h}$, and (d) $C_{6v}$ point groups.}
\label{Fig:VirtualClusters}
\end{figure}

In crystal, crystallographically equivalent atoms are transformed to
each other by combinations of point group symmetry
operation, $\mathcal{R}_i$, non-primitive translation, ${\bm \tau}_j$,
and primitive translation, ${\bm T}_j$, which constitute
symmetry operations of space group.
According to Neumann's principle, the appearance of macroscopic phenomena such as AH and EM effects are determined by crystallographic point group, whose symmetry operations consist only of the rotation part $\mathcal{R}_i$ of the space group.
Therefore, in order to discuss the macroscopic phenomena induced by the antiferromagnetic order, it is necessary to define appropriate order parameters for antiferromagnetic order reflecting the symmetry breaking of the crystallographic point group.

The crystallographic point group corresponding to the space group of crystal is more definitely defined as follows.
The space group $\mathcal{G}$ is decomposed as
\begin{align}
{\mathcal G} = \sum_{i=1}^{N_{\rm coset}}\{\mathcal{R}_{i}|\bm{\tau}_{i}\}{\mathcal H}T,
\label{Eq:cosetSG}
\end{align}
where the subgroup $\mathcal{H}$ consists only of the pure rotation symmetry operations $\{h_1, h_2, \ldots,h_{N_{0}}\}$, $T$ is the group consisting of lattice translations ${\bm T}_{j}$, and $\{\mathcal{R}_{i}|\bm{\tau}_{i}\}$ are representative elements of ${\mathcal G}$ with rotation operation $\mathcal{R}_i$ ($\mathcal{R}_{1}=E$ (identity operation)) and non-primitive translation operation $\bm{\tau}_{i}$ ($\bm{\tau}_{1}=\bm{0}$).
$N_{\rm coset}$ is the number of cosets.
Note that the terms with $i\ge 2$ in Eq.~(\ref{Eq:cosetSG}) exist only in nonsymmorphic space groups.
The crystallographic point group ${\mathcal P}$ corresponding to the space group ${\mathcal G}$ is then defined as
\begin{align}
{\mathcal P} = \sum_{i=1}^{N_{\rm coset}}\{\mathcal{R}_{i}|{\bm 0} \}{\mathcal H}.
\label{Eq:cosetCPG} 
\end{align}

Magnetic configurations in 
an atomic cluster is unambiguously defined by multipoles through Eqs.~(\ref{Eq:CMP}) and (\ref{Eq:CTD}).
It is highly contrast to a direct generation of magnetic structures in crystal, which is not straightforward since the atoms in crystal are related not only by rotation operations but also by translation operations as we discussed in Ref.~\onlinecite{Suzuki2017}.
To avoid difficulties in the direct generation of magnetic structures in crystal, our strategy to generate the symmetry-adapted magnetic structures in crystal is to generate the multipole magnetic configurations at first in a virtual atomic cluster defined under the point group ${\mathcal P}$, and then map the magnetic configurations onto the magnetic structure in crystal so as to preserve the magnetic point group symmetry of the multipole configurations.
The virtual atomic cluster is defined as an atomic cluster consisting of the same number of atoms as the symmetry operations in the crystallographic point group ${\mathcal P}$,
in which their atomic positions are given as the general Wyckoff positions of the corresponding symmorphic space group,
as listed in Table~\ref{Tab:VirtualCluster} for representative point groups.
In the virtual atomic cluster, the magnetic configurations corresponding to the M and MT multipoles, Eqs.~(\ref{Eq:CMP}) and (\ref{Eq:CTD}), are obtained by Eqs.~(\ref{Eq:base_mp}) and (\ref{Eq:base_td}), respectively, with respect to the origin of the virtual atomic cluster.
The examples of the virtual atomic clusters are depicted in Fig.~\ref{Fig:VirtualClusters} for $O_h$, $C_{4h}$, $D_{6h}$, $C_{6v}$ point groups.
Note that the virtual atomic cluster has ambiguity due to the choice of the parameter $x$, $y$, $z$, as seen in the Table~\ref{Tab:VirtualCluster}.
This leads to an arbitrariness of the magnetic configurations in the virtual atomic cluster.
This arbitrariness is largely reduced when the magnetic configuration is mapped onto the atoms at high symmetry sites in crystal, as explained in Sec.~\ref{Sec:Mapping}.

\begin{center}
\begin{table*}[tb]
\caption{List of atomic positions of virtual atomic clusters for representative crystallographic point groups.
Those for other point groups are found in Ref.~\onlinecite{internationaltables2002} as general Wyckoff positions of symmorphic space groups as well as the listed point groups.
}
\begin{tabular}{cl} \hline \hline
       & \multicolumn{1}{c}{atomic positions of virtual atomic clusters} \\
\hline
 $O_h$ & (1) $x,y,z$ (2) $\bar{x}, \bar{y}, z$ (3) $\bar{x},y,\bar{z}$ (4) $x, \bar{y}, \bar{z}$ (5) $z,x,y$ (6)
      $z, \bar{x}, \bar{y}$ (7) $\bar{z}, \bar{x},y$ (8) $\bar{z},x,
      \bar{y}$ \\ 
       & (9) $y, z,x$ (10) $\bar{y}, z, \bar{x}$ (11) $y, \bar{z},
      \bar{x}$ (12) $\bar{y}, \bar{z},x$ (13) $y,x,\bar{z}$ (14)
      $\bar{y}, \bar{x}, \bar{z}$ (15) $y, \bar{x}, z$ (16) $\bar{y},x,z$ \\
      & (17) $x, z,\bar{y}$ (18) $\bar{x}, z,y$ (19) $\bar{x}, \bar{z}, \bar{y}$ (20) $x,\bar{z},y$ (21) $z,y,\bar{x}$ (22) $z,\bar{y},x$ (23) $\bar{z},y,x$ (24) $\bar{z},\bar{y},\bar{x}$ \\
  & (25) $\bar{x},\bar{y},\bar{z}$ (26) $x, y,\bar{z}$ (27) $x,\bar{y},
      z$ (28)$\bar{x},y,z$ (29) $\bar{z},\bar{x},\bar{y}$ (30) $\bar{z},
      x, y$ (31)  $z, x,\bar{y}$ (32) $z,\bar{x},y$ \\
  &   (33) $\bar{y},\bar{z},\bar{x}$ (34)  $y,\bar{z}, x$ (35) $\bar{y}, z, x$ (36) $y, z,\bar{x}$ (37) $\bar{y},\bar{x}, z$ (38) $y, x, z$ (39) $\bar{y}, x,\bar{z}$ (40)  $y,\bar{x},\bar{z}$ \\
 & (41) $\bar{x},\bar{z}, y$ (42)  $x,\bar{z},\bar{y}$ (43) $x, z, y$ (44) $\bar{x}, z,\bar{y}$
(45) $\bar{z},\bar{y}, x$ (46) $\bar{z}, y,\bar{x}$ (47)
      $z,\bar{y},\bar{x}$ (48) $z,y,x$ \\
\hline
 $O$ & (1) $x,y,z$ (2) $\bar{x}, \bar{y}, z$ (3) $\bar{x},y,\bar{z}$ (4) $x, \bar{y}, \bar{z}$ (5) $z,x,y$ (6)
      $z, \bar{x}, \bar{y}$ (7) $\bar{z}, \bar{x},y$ (8) $\bar{z},x,
      \bar{y}$ \\ 
       & (9) $y, z,x$ (10) $\bar{y}, z, \bar{x}$ (11) $y, \bar{z},
      \bar{x}$ (12) $\bar{y}, \bar{z},x$ (13) $y,x,\bar{z}$ (14)
      $\bar{y}, \bar{x}, \bar{z}$ (15) $y, \bar{x}, z$ (16) $\bar{y},x,z$ \\
      & (17) $x, z,\bar{y}$ (18) $\bar{x}, z,y$ (19) $\bar{x}, \bar{z}, \bar{y}$ (20) $x,\bar{z},y$ (21) $z,y,\bar{x}$ (22) $z,\bar{y},x$ (23) $\bar{z},y,x$ (24) $\bar{z},\bar{y},\bar{x}$ \\
\hline
 $T_{d}$  & (1) $x,y, z $ (2) $ \bar{x}, \bar{y}, z $ (3) $ \bar{x},y, \bar{z} $ (4) $ x, \bar{y}, \bar{z}
$ (5) $ z,x,y $ (6) $ z, \bar{x}, \bar{y} $ (7) $ \bar{z}, \bar{x},y $ (8) $ \bar{z},x, \bar{y}$ \\
          & (9) $ y, z,x $ (10) $ \bar{y}, z, \bar{x} $ (11) $ y, \bar{z}, \bar{x} $ (12) $ \bar{y}, \bar{z},x
$ (13) $ y,x, z $ (14) $ \bar{y}, \bar{x}, z $ (15) $ y, \bar{x}, \bar{z} $ (16) $ \bar{y},x, \bar{z}$ \\
          & (17) $ x, z,y $ (18) $ \bar{x}, z, \bar{y} $ (19) $ \bar{x}, \bar{z},y $ (20) $ x, \bar{z}, \bar{y}
$ (21) $ z,y,x $ (22) $ z, \bar{y}, \bar{x} $ (23) $ \bar{z},y, \bar{x}
      $ (24) $ \bar{z}, \bar{y},x $ \\
\hline
 $T_{h}$ & (1) $ x,y, z $ (2) $ \bar{x}, \bar{y}, z $ (3) $ \bar{x},y,\bar{z} $ (4) $ x,\bar{y},\bar{z}
$ (5) $ z,x,y $ (6) $ z,\bar{x},\bar{y} $ (7) $ \bar{z},\bar{x},y $ (8)
      $ \bar{z},x,\bar{y}$ \\
        & (9) $ y,z,x $ (10) $ \bar{y}, z,\bar{x} $ (11) $ y,\bar{z},\bar{x} $ (12) $ \bar{y},\bar{z},x
$ (13) $ \bar{x}, \bar{y}, \bar{z} $ (14) $ x,y,\bar{z} $ (15) $ x,\bar{y}, z $ (16) $ \bar{x},y, z
$ \\
& (17) $ \bar{z}, \bar{x}, \bar{y} $ (18) $ \bar{z},x,y $ (19) $ z,x, \bar{y} $ (20) $ z, \bar{x},y
$ (21) $ \bar{y}, \bar{z}, \bar{x} $ (22) $ y, \bar{z},x $ (23) $ \bar{y}, z,x $ (24) $ y, z,\bar{x}$ \\
\hline
 $C_{4h}$ & (1) $x,y,z$ (2) $\bar{x}, \bar{y}, z$ (3) $\bar{y}, x, z$ (4) $y, \bar{x}, z$
            (5) $\bar{x},\bar{y},\bar{z}$ (6) $x, y, \bar{z}$ (7) $y,
      \bar{x}, \bar{z}$ (8) $\bar{y}, x, \bar{z}$ \\
\hline
 $D_{6h}$ & (1) $x,y, z$ (2) $\bar{y},x-y, z$ (3)$\bar{x}+y, \bar{x}, z$
(4) $\bar{x}, \bar{y}, z$ (5) $y, \bar{x}+y, z$ (6) $x-y,x, z$
(7) $y,x,\bar{z}$ (8) $x-y, \bar{y},\bar{z}$ \\
   &  (9) $\bar{x}, \bar{x}+y, \bar{z}$ (10) $\bar{y}, \bar{x}, \bar{z}$ (11) $\bar{x}+y,y, \bar{z}$ (12) $x,x-y, \bar{z}$ (13) $\bar{x}, \bar{y}, \bar{z}$ (14) $y, \bar{x}+y, \bar{z}$ (15) $x-y, x, \bar{z}$ (16) $x,y,\bar{z}$ \\
   & (17) $\bar{y},z-y,\bar{z}$ (18) $\bar{x}+y, \bar{x}, \bar{z}$ (19)
      $\bar{y},\bar{x}, z$ (20) $\bar{x}+y,y,z $(21) $x,x-y,z$ (22)
      $y,x, z$ (23) $x-y,\bar{y},z$ (24) $\bar{x}, \bar{x}+y,z$ \\
\hline
 $C_{6v}$ & (1) $x,y,z$ (2) $\bar{y}, x-y, z$ (3) $\bar{x}+y,\bar{x}, z$
      (4) $\bar{x}, \bar{y}, z$ (5) $y, \bar{x}+y, z$ (6) $x-y, x, z$
      (7) $\bar{y}, \bar{x}, z$ (8) $\bar{x}+y, y, z$, \\
  &    (9) $x, x-y, z$ (10) $y, x, z$ (11) $x-y, \bar{y}, z$ (12) $\bar{x}, \bar{x}+y, z$ \\
\hline
 $D_{3d}$ & (1) $x,y,z$ (2) $\bar{y},x-y, z$ (3) $\bar{x}+y, \bar{x}, z$
      (4) $\bar{y}, \bar{x}, \bar{z}$ (5) $\bar{x}+y,y,\bar{z}$ (6) $x, x-y,\bar{z}$ (7) $\bar{x}, \bar{y}, \bar{z}$ (8) $ y, \bar{x}+y, \bar{z}$ \\
 &  (9) $ x-y,x, \bar{z} $ (10) $ y,x,z$ (11) $ x-y, \bar{y}, z $ (12) $
      \bar{x}, \bar{x}+y, z$\\
\hline
 \end{tabular}
\label{Tab:VirtualCluster}
\end{table*}
 \end{center}

\subsection{Mapping of multipoles from virtual atomic cluster to crystal}
\label{Sec:Mapping}

The purpose of this section is to obtain the magnetic structures whose
transformation property for the magnetic point group operations is
the same with that of multipole configurations in the virtual atomic cluster.
For this purpose, we first choose one atom in the virtual atomic cluster and one in crystal and set the same M dipole moments on these atoms by setting a mapping from the atoms in virtual atomic cluster to that in crystal.
The whole mapping between the atoms in the virtual atomic cluster and
the symmetrically equivalent atoms in crystal is obtained by identifying
the atoms transformed by the point group symmetry operations, $R_i$, in
the virtual atomic cluster with the atoms transformed by the symmetry
operations of space group, $\{E|\bm{T}_{j}\}\{R_i|\bm{\tau}_i\}h_{k}$, in the crystal for the initially chosen atoms in both systems.
The mapping from the atoms in the virtual atomic cluster to the symmetrically equivalent atoms in crystal obviously does not have one-to-one correspondence.

The M dipole moment on an atom in crystal is obtained by summing up the M dipole moments on the all of the corresponding atoms in the virtual atomic cluster, as will be discussed in Sec.~\ref{Sec:SimpleEx}.
The generated magnetic structures in crystal are fully characterized by the symmetry-adapted multipoles in the virtual atomic cluster through the mapping, which we call the multipole magnetic structures.
We note that the correspondence between generated magnetic structures and multipole configurations depends on the choice of the first mapping between the atoms in virtual atomic cluster and that in crystal.
It occurs especially for low-symmetry crystallographic point groups having the multiple symmetrized bases within the same IREPs and rank.
In this case, we have to specify the representative atoms in the virtual atomic cluster and in crystal to identify the correspondence between the magnetic structures and multipole configurations.
The other arbitrariness for the multipole structures arises from a choice of the parameter ($x$, $y$, $z$) of virtual atomic cluster as listed in Table~\ref{Tab:VirtualCluster}, as already mentioned in Sec.~\ref{Sec:VirtualCluster}.

The advantage to introduce the virtual atomic cluster is that once we generate the multipole magnetic configurations in the virtual atomic cluster, we can systematically generate complete orthonormal basis sets for arbitrary crystal structures by appropriate mappings, in which they share the common 
point group between the atomic cluster and crystal.
This is in a high contrast with our previous scheme proposed in Ref.~\onlinecite{Suzuki2017} to identify the multipoles in crystal.
In the previous scheme, the atomic cluster is defined not in a virtual one but directly for the atoms in the crystal.
 We identified $N_{\rm coset}$ clusters in which one consist of the atoms transformed to each other by the point group operations of $\mathcal{H}$ in Eq.~(\ref{Eq:cosetSG}) and the others are determined from transformation of $\{\mathcal{R}_i|\bm{\tau}_i\}\mathcal{H}$ for the cluster in the crystal unit cell.
The macroscopic multipole moment was obtained by summing up the multipole moments in each cluster
in accordance with a nonsymmorphic space group of Mn$_3$$Z$ ($Z$=Sn, Ge).
However, there are some cases that the higher rank multipoles are ill defined, especially when the magnetic atoms are located at high symmetry sites.
 The present scheme does not cause this problem and is advantageous in its efficiency since the arbitrariness of the generated orthonormal magnetic structures is largely reduced as compared with the conventional method using projection operator~\cite{Bertaut1968, Bertaut1981}, whose generated magnetic structures highly depend on the choice of the trial functions to be operated.
Based on the present procedure, we can systematically and automatically generate magnetic structures corresponding to the multipoles according to the IREPs of the crystallographic point group of the focusing crystal.

\subsection{Orthonormalization of basis set for magnetic structure in crystal}
\label{Sec:Orthonormalization}

The multipole expansion in Eq.~(\ref{Eq:VecPot}) requires infinite number of components to express magnetization distribution in continuous space.
In contrast, the multipole expansion in magnetic structures that does not break the crystal periodicity is represented by a linear combination of 3$N_{\rm atom}$ orthogonal basis set, where $N_{\rm atom}$ is the number of atoms in a crystal unit cell.
In this subsection, we explain how to obtain the 3$N_{\rm atom}$ orthonormal basis set which is sufficient to express a uniform magnetic order.

For notational convenience, we introduce the vector notation for uniform magnetic structures, $\{{\bm a}\}=({\bm a}_1,{\bm a}_2,\cdots,{\bm a}_{N_{\rm atom}})$, where ${\bm a}_{i}$ represents the three component vector on the $i$-th atom in the crystal unit cell.
Usually, the magnetic structure $\{{\bm m}_i\}$ is decomposed into the ferromagnetic part $\{{\bm m}_i^{\rm FM}\}$, where ${\bm m}_i^{\rm FM}=\sum_i{\bm m}_i/N_{\rm atom}$, and the rest antiferromagnetic part $\{{\bm m}_{i}^{\rm AFM}\}$, where ${\bm m}_{i}^{\rm AFM}={\bm m}_{i}-{\bm m}_i^{\rm FM}$\cite{Ederer2007}.
Such a decomposition for the magnetic structure is generalized to obtain the 3$N_{\rm atom}$ orthonormal complete basis set by using Gram-Schmidt orthonormalization procedure as follows:
\begin{align}
\{{\bm e}_{1\gamma}^{1}\}&\equiv \frac{\{{\bm u}^{1}_{1\gamma}\}}{\sqrt{(\{{\bm u}^{1}_{1\gamma}\}\cdot\{{\bm u}^{1}_{1\gamma}\})}},
\label{Eq:GS_initial}\\
\{{\bm v}_{\ell\gamma}^{\mu}\} &=\{{\bm u}_{\ell\gamma}^{\mu}\}-\sum_{\ell'=1}^{\ell}\sum_{\mu'=1}^{\mu}\sum_{\gamma'=1}^{\gamma-1}(\{{\bm u}_{\ell\gamma}^{\mu'}\}\cdot\{{\bm e}_{\ell'\gamma'}^{\mu'}\})\{{\bm e}_{\ell'\gamma'}^{\mu'}\},
\label{Eq:GS_Orthogonal}\\
\{{\bm e}_{\ell\gamma}^{\mu}\}&=\frac{\{{\bm v}_{\ell\gamma}^{\mu}\}}{\sqrt{(\{{\bm v}_{\ell\gamma}^{\mu}\}\cdot\{{\bm v}_{\ell\gamma}^{\mu}\})}},
\label{Eq:GS_Orthonormal}
\end{align}
where $\mu=1$ and $2$ represent the M and MT multipoles, respectively.
The initial ${\bm u}^{1}_{1\gamma}$ in Eq.~(\ref{Eq:GS_initial}) is set as the M dipole moments, $M_{x}$, $M_{y}$, and $M_{z}$ for $\gamma=1,2,3$, respectively.
The iterated calculation of the Gram-Schmidt orthonormalization procedure, Eqs.~(\ref{Eq:GS_Orthogonal}) and (\ref{Eq:GS_Orthonormal}), starting from the lower-rank multipoles to higher ones as shown in Fig.~\ref{Fig:GenerationFlow}, automatically generates orthonormal complete basis set of the uniform magnetic structures classified according to the IREPs of crystallographic point group.
After orthonormalization procedure, the magnetic structures corresponding to the higher-rank multipole may not be the pure multipole with definite rank as in Eq.~(\ref{Eq:GS_Orthogonal}).
However, since the subtraction in Eq.~(\ref{Eq:GS_Orthogonal}) is to eliminate the overlap between the highest-rank multipole and the lower ones, and $\{{\bm e}_{\ell\gamma}^{\mu}\}$ in Eq.~(\ref{Eq:GS_Orthonormal}) always contains $\{{\bm u}_{\ell\gamma}^{\mu}\}$, it can be regarded as the rank-$\ell$ multipole.

The generated finite norm 
of vectors $\{{\bm e}_{\ell\gamma}^{\mu}\}$ are stored sequentially as $\{{\bm e}^{i}\}$ with the sequential indices $i=1,...,3N_{\rm atom}$.
With this procedure, the ferromagnetic structures $\{{\bm e}^{i}\}$ with $i\le 3$ are orthogonal to the antiferromagnetic structures with $i>3$.
The obtained magnetic structure basis set are orthonormal, i.e., $(\{{\bm e}^{i}\}\cdot \{{\bm e}^{j}\})=\delta_{ij}$.
Since the obtained basis set is complete for the uniform magnetic structures, arbitrary uniform magnetic structure can be expressed as $\{{\bm m}\}=\sum_{i=1}^{3N_{\rm atom}}c_{i}\{{\bm e}^{i}\}$ with $c_{i}=(\{{\bm m}\}\cdot \{{\bm e}^{i}\})$.
Note that the relation $\sum_{i}^{3N_{\rm atom}}|c_{i}|^2=\sum_{i}^{N_{\rm atom}}|{\bm m}_{i}|^2$ holds.

\section{Examples of multipole expansion for crystals}
\subsection{Simple examples}
\label{Sec:SimpleEx}

\begin{figure}[tb]
\includegraphics[width=0.8\linewidth]{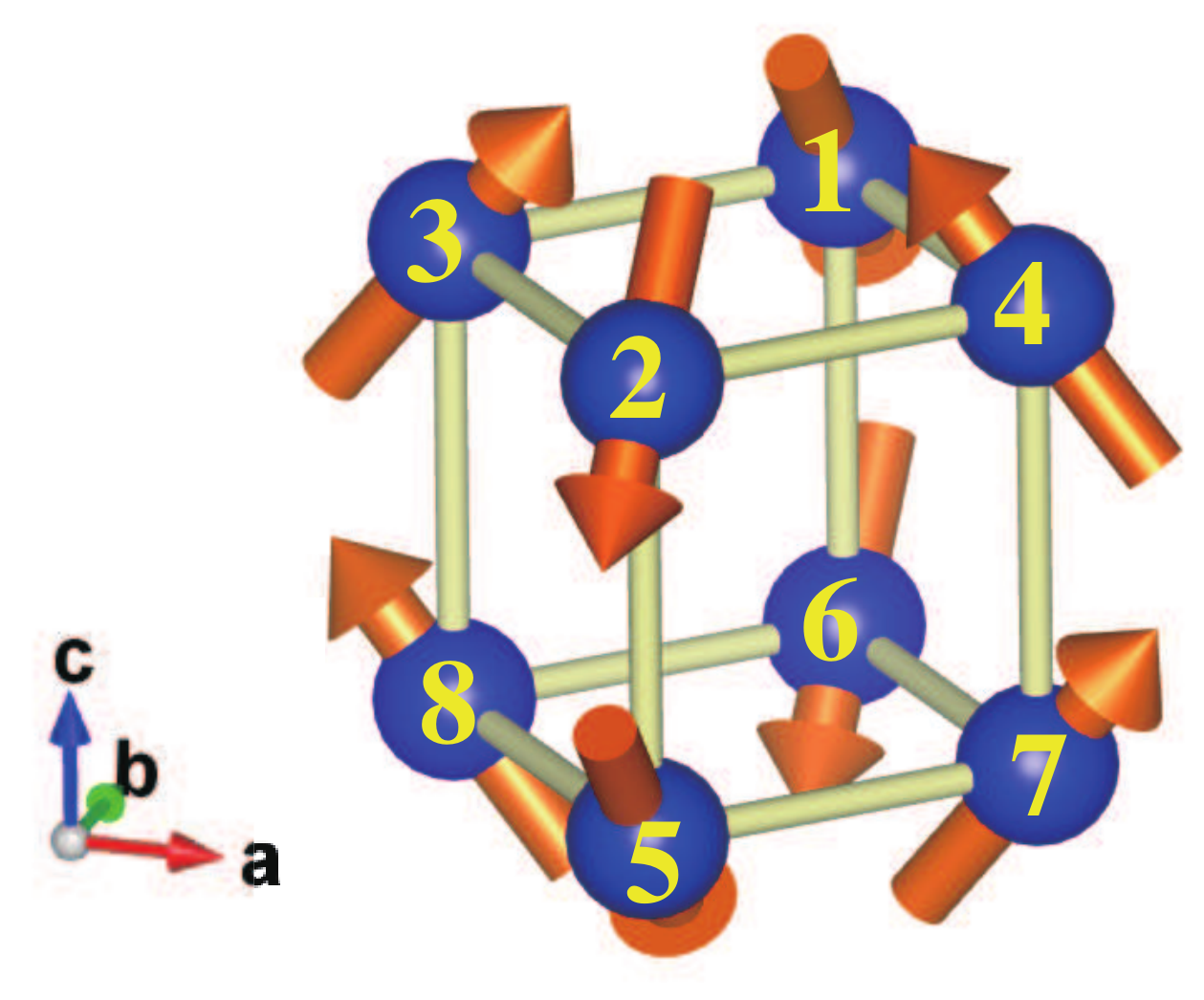}
\caption{
Magnetic configuration of the MT quadrupole in $B_{g}$ IREP of $C_{4h}$ point group in the virtual atomic cluster, placing an atom labeled 1 at (0,0.3,0.3), which is obtained by Eq.~(\ref{Eq:base_td}) with $\mathcal{Y}_{25}$ in Eq.~(\ref{Eq:symY25}).}
\label{Fig:R2Toroid_VC}
\end{figure}

\begin{figure}[tb]
\includegraphics[width=0.9\linewidth]{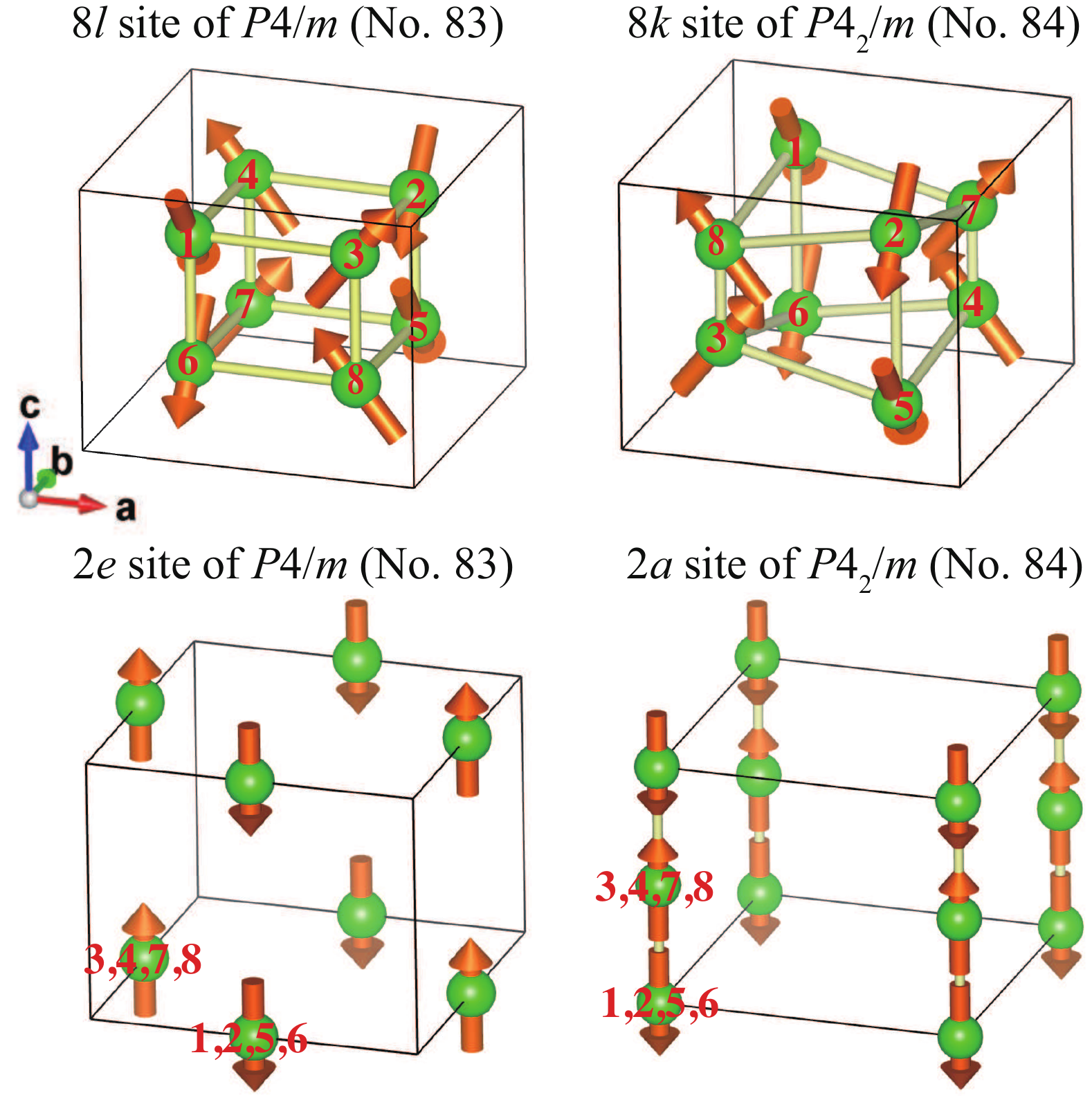}
\caption{MT quadrupoles in crystals belonging to $P4/m$ and $P4_{2}/m$ space groups, which are obtained by mapping from the $C_{4h}$ virtual atomic cluster as shown in Fig.~\ref{Fig:R2Toroid_VC}.}
\label{Fig:R2Toroid_Cryst}
\end{figure}

We here illustrate the correspondence between the symmetry-adapted multipole magnetic configurations in the virtual atomic cluster and magnetic structures in crystal.
We take two simple examples, i.e., the atoms placed at 8$l$ and 2$e$
Wyckoff sites in symmorphic space group $P4/m$ (No.~83), and 8$k$ and
2$a$ Wyckoff sites in nonsymmorphic space group $P4_{2}/m$ (No.~84).
The crystallographic point group of both space groups is $C_{4h}$, which corresponds to a virtual atomic cluster consisting of eight atoms transformed to each other through the symmetry operations 
as shown in Fig.~\ref{Fig:VirtualClusters} (b).
In Fig.~\ref{Fig:R2Toroid_VC}, two fold rotation $C_{2}$ transforms the atom 1 to atom 2, four fold rotation $C_{4}$ to atom 3, $C_{4}^{-1}$ to atom 4, space inversion $I$ to atom 5, $IC_{2}$ to atom 6, $IC_{4}$ to atom  7, and $IC_{4}^{-1}$ to atom 8.
These transformation relations provide mapping from the atoms in the virtual atomic cluster to symmetrically equivalent atoms in the crystal as discussed in Sec.~\ref{Sec:Mapping}.
Figures~\ref{Fig:R2Toroid_VC} and \ref{Fig:R2Toroid_Cryst} show the relation between the atoms in the virtual atomic cluster and the atoms at the Wyckoff sites in crystal.


For the $C_{4h}$ point group, the rank-2 spherical harmonics are symmetrized as
\begin{align}
\mathcal{Y}_{21} &= Y_{20}\quad (A_g), \cr
\mathcal{Y}_{22} &= \frac{1}{\sqrt{2}}(Y_{2-1}-Y_{21})\quad (E_g), \cr
\mathcal{Y}_{23} &= \frac{1}{\sqrt{2}i}(Y_{2-1}+Y_{21})\quad (E_g),
\label{Eq:symY25} \\
\mathcal{Y}_{24} &= \frac{1}{\sqrt{2}}(Y_{2-2}+Y_{22})\quad (B_g), \cr
\mathcal{Y}_{25} &= \frac{1}{\sqrt{2}i}(Y_{2-2}-Y_{22})\quad (B_g). \nonumber
\end{align}
The multipole configurations are calculated from Eq.~(\ref{Eq:base_mp}) and (\ref{Eq:base_td}) on the virtual atomic cluster as shown in Fig.~\ref{Fig:VirtualClusters}(b), where the atom 1 is placed at (0,0.3,0,3).
Figure~\ref{Fig:R2Toroid_VC} shows the magnetic configuration generated from Eq.~(\ref{Eq:base_td}) with the symmetrized spherical harmonics $\mathcal{Y}_{25}$ in Eq.~(\ref{Eq:symY25}), corresponding to the MT quadrupole, $T_{xy}$, classified to $B_{g}$ IREP of $C_{4h}$ point group~\cite{Hayami2018}.
This magnetic configuration in the virtual atomic cluster is mapped to symmetrically equivalent atoms, denoted by the signatures of Wyckoff positions, to obtain the corresponding magnetic structures in crystals as shown in Fig.~\ref{Fig:R2Toroid_Cryst}.
For the general Wyckoff positions in each crystal ($8l$ site of
$P4/m$ and $8k$ site of $P4_{2}/m$), the magnetic structures are straightforwardly obtained through the mapping (Fig.~\ref{Fig:R2Toroid_Cryst}).
Meanwhile, for $2e$ site of $P4/m$ and $2a$ site of $P4_{2}/m$, multiple atoms in the virtual cluster are mapped to the same atoms in crystal, leading to the cancellation of the M dipole moment that is consistent with the symmetry in these Wyckoff sites.

\subsection{Pyrochlore structure}
\begin{table}[tb]
\caption{
 Relation between the multipoles corresponding to the orthonormal magnetic structures for pyrochlore crystal structure and IREP for the crystallographic point group $O_{h}$, as well as the magnetic point group with its principal axis.
The active component of the AH conductivity (AHC) is also shown.
}
\begin{tabular}{cccccc} \hline\hline
 No. & IREP & multipole & MPG & P. axis & AHC \\
\hline
 1   &  $T_{1g}$  & $M_{x}$          & $4/mm'm'$    & [100] & $\sigma_{yz}$ \\
 2   &            & $M_{y}$          & $4/mm'm'$    & [010] & $\sigma_{zx}$ \\
 3   &            & $M_{z}$          & $4/mm'm'$    & [001] & $\sigma_{xy}$ \\ \hline
 4   &  $E_{g}$   & $T_{v}$          & $4/mmm$      & [001] & --- \\
 5   &            & $T_{u}$          & $4'/mmm'$    & [001] & --- \\ \hline
 6   &  $T_{2g}$  & $T_{yz}$         & $4'/mm'm$    & [100] & ---\footnotemark[1] \\
 7   &            & $T_{zx}$         & $4'/mm'm$    & [010] & ---\footnotemark[1] \\
 8   &            & $T_{xy}$         & $4'/mm'm$    & [001] & ---\footnotemark[1] \\ \hline
 9   &  $A_{2g}$  & $M_{xyz}$        & $m\bar{3}m'$ & [001] & --- \\ \hline
10   &  $T_{1g}$  & $M_{x}^{\alpha}$ & $4/mm'm'$    & [100] & $\sigma_{yz}$ \\
11   &            & $M_{y}^{\alpha}$ & $4/mm'm'$    & [010] & $\sigma_{zx}$ \\
12   &            & $M_{z}^{\alpha}$ & $4/mm'm'$    & [001] & $\sigma_{xy}$ \\
\hline
\end{tabular}
\footnotetext[1]{The multipole magnetic structures characterized by one of $T_{yz}$, $T_{zx}$, and $T_{xy}$ does not induce the AH effect, but those obtained by linear combinations of these MT quadrupoles can induce the AH conductivity in general due to the absence of rotation symmetries.}
\label{Tab:Pyrochlore_CMP_sym}
\end{table}

\begin{figure}[tb]
\includegraphics[width=0.9\linewidth]{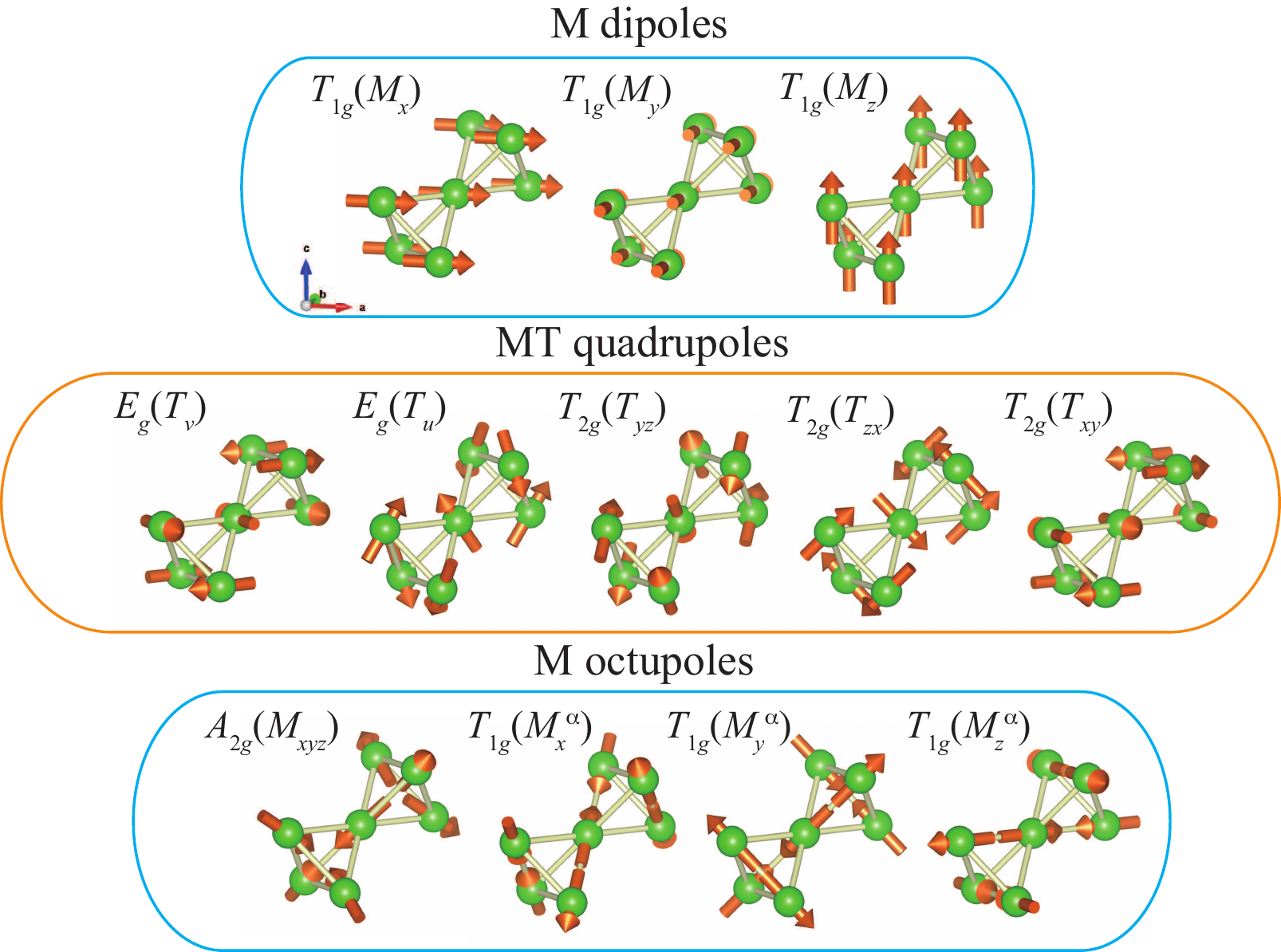}
\caption{Multipole magnetic structures in pyrochlore structure characterized by the IREP of $O_h$ point group and its rank.
Notations of multipoles are adapted from Tables in Ref.~\onlinecite{Hayami2018}.}
\label{Fig:Pyrochlore}
\end{figure}

Pyrochlore crystal structure belongs to the space group $Fd\bar{3}m$ (No.~227), whose crystallographic point group $O_h$ leads to the virtual atomic cluster as shown in Fig.~\ref{Fig:VirtualClusters}(a).
The generation procedure of orthonormal magnetic structures gives twelve orthonormal bases characterized by the M dipole and octupole, and MT quadrupole as shown in Table~\ref{Tab:Pyrochlore_CMP_sym}.
In the pyrochlore structure, the inversion symmetry breaking is accompanied by the breaking of the translation symmetries since the two atoms related by the space inversion are also transformed by the commensurate translation of the crystal.
Therefore, a uniform magnetic structure 
must preserve the space inversion symmetry, restricting the parity-even multipole structures without EM effects.
The generated magnetic structure coincides with the basis set obtained in earlier work.~\cite{Wills2006}.

\begin{figure}[tb]
\includegraphics[width=0.9\linewidth]{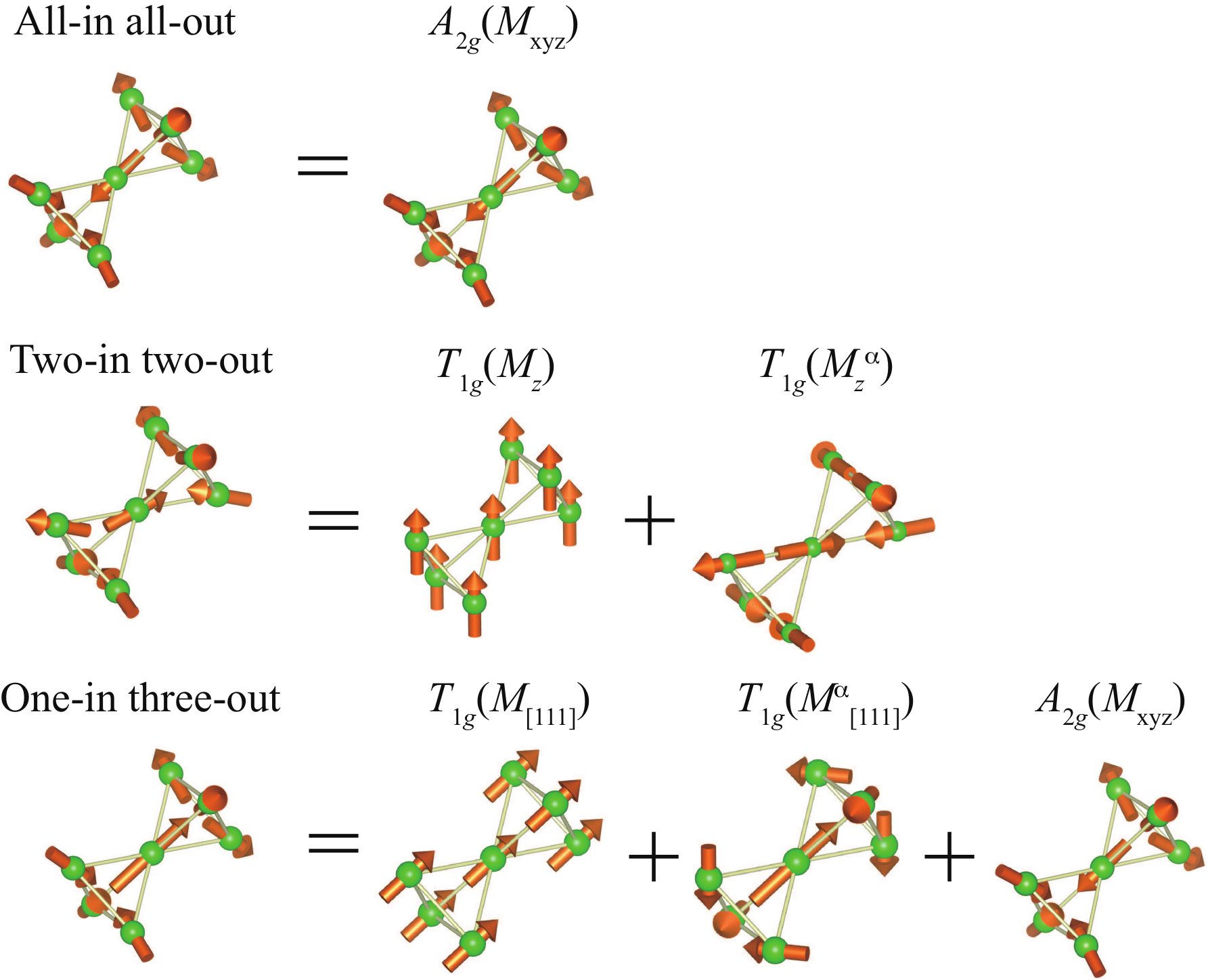}
\caption{The multipole expansion of all-in all-out, two-in two-out and one-in three-out magnetic structures in pyrochlore crystal structure.
}
\label{Fig:MPexpansion_Pyrochlore}
\end{figure}

The pyrochlore compounds are known to show a variety of magnetic order depending on the atomic constitutions, such as all-in all-out, two-in two-out, and one-in three-out magnetic orders.
The relation between the observed magnetic structures and the M multipoles are shown in Table~\ref{Tab:RealMaterials}.
The corresponding uniform magnetic orders are shown in Fig.~\ref{Fig:Pyrochlore}.
All-in all-out magnetic structures, reported in Cd$_2$Os$_2$O$_7$ and Er$_2$Ti$_2$O$_7$, correspond to $M_{xyz}$-octupole in Fig.~\ref{Fig:Pyrochlore}.
The magnetic point group of the all-in all-out structure prohibits to induce AH effect.
Meanwhile, two-in two-out magnetic structure is expressed by a linear combination of the M dipole and octupole belonging to $T_{1g}$ IREP as shown in Fig.~\ref{Fig:MPexpansion_Pyrochlore}.
The two-in two-out magnetic structure is allowed to induce the AH effect by the magnetic point group symmetry.
Since arbitrary linear combination does not change the symmetry, pure $T_{1g}$ octupole structure $M_{z}^{\alpha}$ without net magnetization can also induce the AH effect.
This fact may be relevant to the mechanism of the AH effect of Pr$_2$Ir$_2$O$_7$ without net magnetization~\cite{Machida2010,Nakatsuji2011}.
The order parameter for the magnetic structures of Yb$_2$Ti$_2$O$_7$, Yb$_2$Sn$_2$O$_7$, and Tm$_2$Mn$_2$O$_7$ are also characterized by the $T_{1g}$ multipoles, but in these cases, the M dipole moment is dominant~\cite{Gaudet2016,Yaouanc2013,Pomjakushina2015}.

\begin{center}
\begin{table*}[tb]
\caption{Relation between the experimentally observed magnetic structures and the M multipoles in pyrochlore and hexagonal $AB$O$_3$ crystal structures.
$M_{[111]}=(M_{x}+M_{y}+M_{z})/\sqrt{3}$ and $M_{[111]}^{\alpha}$ is similarly defined.
}
\begin{tabular}{cccc} \hline\hline
Multipole & IREP & Name & Materials \\ \hline\hline
Pyrochlore \\ \hline
$M_{xyz}$ & $A_{2g}$ & all-in all-out & Cd$_2$Os$_2$O$_7$~\cite{Yamaura2012}, Er$_2$Ti$_2$O$_7$~\cite{Poole2007} \\
$T_{xy}$ & $T_{2g}$ & --- & Gd$_2$Sn$_2$O$_7$~\cite{Wills2006}, Er$_2$Ru$_2$O$_7$~\cite{Taira2003} \\
($M_{z}$, $M_{z}^{\alpha}$) & $T_{1g}$ & 2-in 2-out & Ho$_2$Ru$_2$O$_7$\cite{Wiebe2004}, Tb$_2$Sn$_2$O$_7$\cite{Mirebeau2005} \\
$M_{z}^{\alpha}=2\sqrt{2}M_{z}$ & & & \\
($M_{z}$, $M_{z}^{\alpha}$) & $T_{1g}$ & --- & Yb$_{2}$Ti$_{2}$O$_{7}$\cite{Gaudet2016}, Yb$_{2}$Sn$_{2}$O$_{7}$~\cite{Yaouanc2013} \\
$M_{z}$ dominant & & & Tm$_{2}$Mn$_{2}$O$_{7}$\cite{Pomjakushina2015} \\
($M_{[111]}$, $M_{[111]}^{\alpha}$, $M_{xyz}$) & $T_{1g}$ & 1-in 3-out & Tb$_2$Ti$_2$O$_7$\footnotemark[2]~\cite{Sazonov2013} \\ \hline\hline
Hexagonal $AB$O$_3$ \\ \hline
$M_{u}$ & $A_{2}$ & --- & LuFeO$_3$~\cite{Disseler2015}, ScMnO$_3$ (75-129K)~\cite{Munoz2000} \\
$M_{3b}$ & $B_{1}$ & --- & HoMnO$_3$ (below 40K)~\cite{Brown2006} \\
$M_{3a}$ & $B_{2}$ & --- & HoMnO$_3$ (40-75K)~\cite{Brown2006}, YbMnO$_3$~\cite{Fabreges2008} \\
($M_{u}$, $T_{z}$) & $A_{2}\oplus A_{1}$ & --- & ScMnO$_3$ (below 75K)~\cite{Munoz2000} \\
($M_{3b}$, $M_{3a}$) & $B_{1}\oplus B_{2}$ & --- & YMnO$_3$~\cite{Brown2006} \\
\hline
\end{tabular} 
\footnotetext[2]{Stabilized under magnetic fields above 5T.}
\label{Tab:RealMaterials}
\end{table*}
\end{center}

\subsection{Hexagonal $AB$O$_3$}
\label{Sec:HexABO3}
\begin{table}[tb]
\caption{
 Relation between the multipoles corresponding to the orthonormal magnetic structures for hexagonal $AB$O$_3$ and IREP of $C_{6v}$, as well as the magnetic point group (MPG) with its principal axis. The active components of the AH conductivity (AHC) and EM tensor are also shown.
}
\begin{tabular}{ccccccc} \hline\hline
 No.   & IREP& multipole & MPG & P. axis & AHC   & EM \\
\hline
 1     & $E_{1}$     & $M_{x}$         & $mm'2'$ &   [100]     & $\sigma_{yz}$ & $\alpha_{xz}$, $\alpha_{zx}$ \\
 2     &             & $M_{y}$         & $m'm2'$ &   [100]     & $\sigma_{zx}$ & $\alpha_{yz}$, $\alpha_{zy}$ \\ \hline
 3     & $A_{2}$     & $M_{z}$         & $6m'm'$ &   [001]     & $\sigma_{xy}$ & $\alpha_{xx}=\alpha_{yy}$,$\alpha_{zz}$ \\ \hline
 4     & $E_{1}$     & $T_{x}$         & $m'm2'$ &   [100]     & $\sigma_{zx}$ & $\alpha_{yz}$, $\alpha_{zy}$ \\
 5     &             & $T_{y}$         & $mm'2'$ &   [100]     & $\sigma_{yz}$ & $\alpha_{xz}$, $\alpha_{zx}$ \\ \hline
 6     & $A_{1}$     & $T_{z}$         & $6mm$   &   [001]     & --- & $\alpha_{xy}=-\alpha_{yx}$ \\ \hline
 7     & $A_{2}$     & $M_{u}$         & $6m'm'$ &   [001]     & $\sigma_{xy}$ & $\alpha_{xx}=\alpha_{yy},\alpha_{zz}$ \\ \hline
 8     & $E_{2}$     & $M_{v}$   & $m'm'2$ &   [100]     & $\sigma_{xy}$ & $\alpha_{xx}$,$\alpha_{yy}$,$\alpha_{zz}$ \\
 9     &             & $M_{xy}$        & $mm2$   &   [100]     & --- & $\alpha_{xy}$, $\alpha_{yx}$ \\ \hline
10     & $E_{1}$     & $T_{yz}$        & $mm'2'$ &   [100]     & $\sigma_{yz}$ & $\alpha_{xz}$, $\alpha_{zx}$ \\
11     &             & $T_{zx}$        & $m'm2'$ &   [100]     & $\sigma_{zx}$ & $\alpha_{yz}$, $\alpha_{zy}$ \\ \hline
12     & $E_{2}$     & $T_{v}$   & $mm2$   &   [100]     & --- & $\alpha_{xy}$, $\alpha_{yx}$ \\
13     &             & $T_{xy}$        & $m'm'2$ &   [100]     & $\sigma_{xy}$ & $\alpha_{xx}$,$\alpha_{yy}$,$\alpha_{zz}$\\ \hline
14     & $B_{1}$     & $M_{3b}$ & $6'mm'$ &   [001]     & --- & --- \\ \hline
15     & $B_{2}$     & $M_{3a}$ & $6'm'm$ &   [001]     & --- & --- \\ \hline
16     & $B_{1}$     & $T_{3a}$ & $6'mm'$ &   [001]     & --- & --- \\ \hline
17     & $E_{2}$     & $T_{xyz}$       & $m'm'2$ &   [100]     & $\sigma_{xy}$ & $\alpha_{xx}$,$\alpha_{yy}$,$\alpha_{zz}$ \\
18     &             & $T_{z}^{\beta}$ & $mm2$   &   [100]     & --- & $\alpha_{xy}$, $\alpha_{yx}$ \\
\hline
\end{tabular}
\label{Tab:HexagonalABO3_CMP_sym}
\end{table}

\begin{center}
\begin{figure*}[tb]
\includegraphics[width=0.9\linewidth]{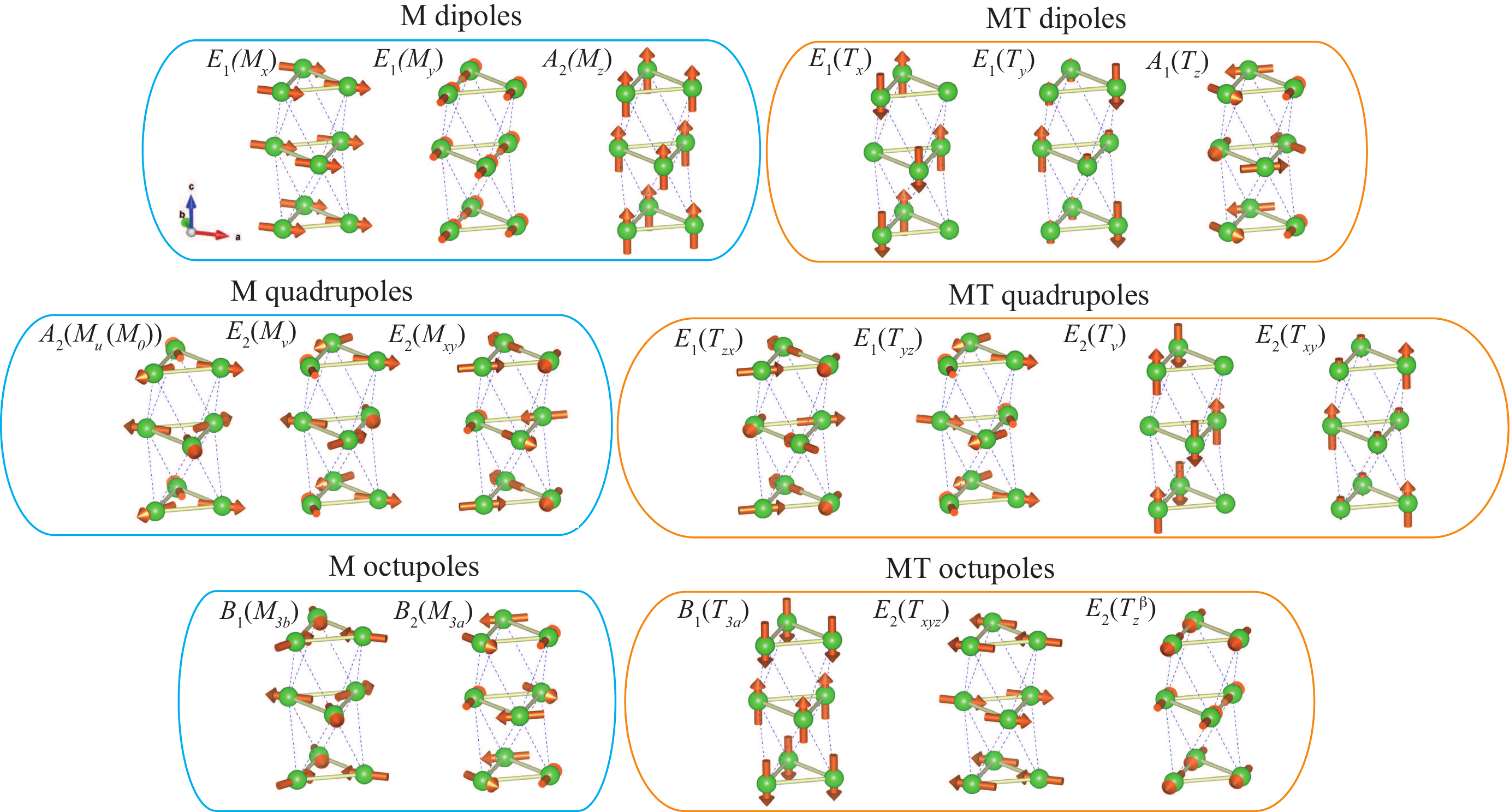}
\caption{Multipole magnetic structures in $AB$O$_3$ characterized by the IREP of $C_{6v}$ point group and its rank.
Notations of multipoles are adapted from Tables in Ref.~\onlinecite{Hayami2018}.}
\label{Fig:ABO3_multipole}
\end{figure*}
\end{center}

Hexagonal $AB$O$_3$ compounds belong to $P6_{3}cm$ (No.~185) space group, whose crystallographic point group is $C_{6v}$, and the corresponding virtual cluster is 
shown in Fig.~\ref{Fig:VirtualClusters}(d).
The magnetic structure generation gives eighteen magnetic structures as shown in Fig.~\ref{Fig:ABO3_multipole}.
The present scheme generates the real basis set, while it contains the complex expressions for two dimensional IREPs, $E_{1}$ and $E_{2}$, in the previous work~\cite{Munoz2000}.
In Table~\ref{Tab:HexagonalABO3_CMP_sym}, we summarize the relation between the multipole magnetic structures and possible AH and EM effects induced under uniform magnetic order.
Table~\ref{Tab:HexagonalABO3_CMP_sym} shows that the higher-rank multipoles are necessary to fully describe the EM effects.

The magnetic order characterized by $M_{u}$ quadrupole is recognized in LuFeO$_3$, and in a temperature range of  75K-129K in ScMnO$_3$~\cite{Munoz2000}.
The $M_{u}$ quadrupole induces the diagonal components in the EM tensor, $\alpha_{xx}=\alpha_{yy}$ and $\alpha_{zz}$.
The M quadrupole belongs to $A_2$ IREP, which is the same as that of $M_{z}$ dipole in the $C_{6v}$ crystallographic point group, and hence it can also induce the AH conductivity $\sigma_{xy}$ as shown in Table~\ref{Tab:RealMaterials}.
We note that the {$M_{u}$} quadrupole structure is also regarded as the monopole magnetic structure characterized by $M_{0}$ since the same magnetic structure is obtained by using Eq.~(\ref{Eq:base_mono}) with the appropriate mapping from the virtual cluster to the crystal unit cell.

The magnetic structures observed in HoMnO$_3$~\cite{Brown2006} and YbMnO$_3$~\cite{Fabreges2008} are characterized by $M_{3a}$ and $M_{3b}$ octupoles, which do not induce the EM effect (Table~\ref{Tab:HexagonalABO3_CMP_sym}).
The magnetic structures in YMnO$_3$ and the low-temperature phase in ScMnO$_3$ in Table~\ref{Tab:RealMaterials} are characterized by linear combinations of the multipole structures with different IREPs, which lowers the magnetic point group symmetry, and may lead to additional finite components in the AH conductivity and EM {\rm tensors.}
For instance, the ($M_u$, $T_z$) multipole structure of low-temperature phase of ScMnO$_3$ only preserves 6 fold rotation symmetry along $z$-axis and can have finite $\sigma_{xy}$ for the AH conductivity~\cite{Kleiner1966,Seemann2015,Suzuki2017}, and finite $\alpha_{xx}=\alpha_{yy}$, $\alpha_{zz}$, and $\alpha_{xy}=-\alpha_{yx}$ for the EM coefficients~\cite{Birss1964,Rivera2009}.
Meanwhile, the ($M_{3b}$, $M_{3a}$) multipole structure of YMnO$_3$ preserves the magnetic point group symmetry operation of 6 fold rotation along $z$-axis combined with the time reversal operation, and no AH and EM effects are expected.

\section{Summary}
We have proposed a scheme to efficiently generate the symmetry-adapted orthonormal magnetic structures in crystallographic point group by introducing a virtual atomic cluster to perform the multipole expansion.
With this method, we can obtain magnetic structure that is fully characterized by the M and MT multipole as the suitable order parameters.
We have introduced a virtual atomic cluster to obtain the magnetic structures preserving the magnetic point group symmetry of the multipoles.
We have applied the proposed method to pyrochlore and hexagonal $AB$O$_3$ crystal structures.
For the pyrochlore crystal structure, we have investigated all-in all-out, two-in two-out, and one-in three-out magnetic structures and found that the two-in two-out and one-in three-out magnetic structures are able to be transformed continuously to the pure antiferromagnetic (octupole) structures without net magnetization, leading to the AH effect.
For the hexagonal $AB$O$_3$ crystal structure, the expression of the EM effect is fully identified by the multipole magnetic structures.
The proposed scheme paves a way to generate the multipole magnetic structures in compatible with the crystallographic point group, which is essential for macroscopic phenomena following Neumann's principle, and is useful to search for desired functional magnetic materials.

\section*{Acknowledgments}
This work was supported by JSPS KAKENHI Grant Numbers JP15K17713 (MTS),
JP15H05883 (J-Physics) (MTS), JP18H04230 (MTS), JP16H04021 (MTS), JP15K05176(HK),
JP15H05885 (J-Physics) (HK), JP16H06590(SH), JP18H04296
(J-Physics) (SH), JP18K13488 (SH), JP16H06345 (RA), JST
PREST (MTS), CREST Grant Number JPMJCR15Q5 (RA\& MTS).
T. N. is supported by RIKEN Special Postdoctoral Researchers Program.

\bibliographystyle{apsrev}
\bibliography{CMPTheory_MagnGen}

\begin{thebibliography}{44}
\expandafter\ifx\csname natexlab\endcsname\relax\def\natexlab#1{#1}\fi
\expandafter\ifx\csname bibnamefont\endcsname\relax
  \def\bibnamefont#1{#1}\fi
\expandafter\ifx\csname bibfnamefont\endcsname\relax
  \def\bibfnamefont#1{#1}\fi
\expandafter\ifx\csname citenamefont\endcsname\relax
  \def\citenamefont#1{#1}\fi
\expandafter\ifx\csname url\endcsname\relax
  \def\url#1{\texttt{#1}}\fi
\expandafter\ifx\csname urlprefix\endcsname\relax\def\urlprefix{URL }\fi
\providecommand{\bibinfo}[2]{#2}
\providecommand{\eprint}[2][]{\url{#2}}

\bibitem[{\citenamefont{Birss}(1962)}]{Birss1962}
\bibinfo{author}{\bibfnamefont{R.~R.} \bibnamefont{Birss}},
  \bibinfo{journal}{Proc. Phys. Soc.} \textbf{\bibinfo{volume}{79}},
  \bibinfo{pages}{946} (\bibinfo{year}{1962}).

\bibitem[{\citenamefont{Birss}(1964)}]{Birss1964}
\bibinfo{author}{\bibfnamefont{R.~R.} \bibnamefont{Birss}},
  \emph{\bibinfo{title}{Symmetry and magnetism}}
  (\bibinfo{publisher}{North-Holland Pub. Co.}, \bibinfo{year}{1964}).

\bibitem[{\citenamefont{Kleiner}(1966)}]{Kleiner1966}
\bibinfo{author}{\bibfnamefont{W.}~\bibnamefont{Kleiner}},
  \bibinfo{journal}{Phys. Rev.} \textbf{\bibinfo{volume}{142}},
  \bibinfo{pages}{318} (\bibinfo{year}{1966}).

\bibitem[{\citenamefont{Rivera}(2009)}]{Rivera2009}
\bibinfo{author}{\bibfnamefont{J.-P.} \bibnamefont{Rivera}},
  \bibinfo{journal}{Eur. Phys. J. B} \textbf{\bibinfo{volume}{71}},
  \bibinfo{pages}{299} (\bibinfo{year}{2009}).

\bibitem[{\citenamefont{Szaller et~al.}(2013)\citenamefont{Szaller,
  Bord{\'a}cs, and K{\'e}zsm{\'a}rki}}]{Szaller2013}
\bibinfo{author}{\bibfnamefont{D.}~\bibnamefont{Szaller}},
  \bibinfo{author}{\bibfnamefont{S.}~\bibnamefont{Bord{\'a}cs}},
  \bibnamefont{and}
  \bibinfo{author}{\bibfnamefont{I.}~\bibnamefont{K{\'e}zsm{\'a}rki}},
  \bibinfo{journal}{Phys. Rev. B} \textbf{\bibinfo{volume}{87}},
  \bibinfo{pages}{014421} (\bibinfo{year}{2013}).

\bibitem[{\citenamefont{Seemann et~al.}(2015)\citenamefont{Seemann,
  K\"{o}dderitzsch, Wimmer, and Ebert}}]{Seemann2015}
\bibinfo{author}{\bibfnamefont{M.}~\bibnamefont{Seemann}},
  \bibinfo{author}{\bibfnamefont{D.}~\bibnamefont{K\"{o}dderitzsch}},
  \bibinfo{author}{\bibfnamefont{S.}~\bibnamefont{Wimmer}}, \bibnamefont{and}
  \bibinfo{author}{\bibfnamefont{H.}~\bibnamefont{Ebert}},
  \bibinfo{journal}{Phys. Rev. B} \textbf{\bibinfo{volume}{92}},
  \bibinfo{pages}{155138} (\bibinfo{year}{2015}).

\bibitem[{\citenamefont{Suzuki et~al.}(2017)\citenamefont{Suzuki, Koretsune,
  Ochi, and Arita}}]{Suzuki2017}
\bibinfo{author}{\bibfnamefont{M.-T.} \bibnamefont{Suzuki}},
  \bibinfo{author}{\bibfnamefont{T.}~\bibnamefont{Koretsune}},
  \bibinfo{author}{\bibfnamefont{M.}~\bibnamefont{Ochi}}, \bibnamefont{and}
  \bibinfo{author}{\bibfnamefont{R.}~\bibnamefont{Arita}},
  \bibinfo{journal}{Phys. Rev. B} \textbf{\bibinfo{volume}{95}},
  \bibinfo{pages}{094406} (\bibinfo{year}{2017}).

\bibitem[{\citenamefont{Shindou and Nagaosa}(2001)}]{Shindou2001}
\bibinfo{author}{\bibfnamefont{R.}~\bibnamefont{Shindou}} \bibnamefont{and}
  \bibinfo{author}{\bibfnamefont{N.}~\bibnamefont{Nagaosa}},
  \bibinfo{journal}{Phys. Rev. Lett.} \textbf{\bibinfo{volume}{87}},
  \bibinfo{pages}{116801} (\bibinfo{year}{2001}).

\bibitem[{\citenamefont{{\v S}mejkal et~al.}(2019)\citenamefont{{\v S}mejkal,
  Gonz{\' a}lez-Hern{\' a}ndez, and Sinova}}]{Smejkal2019}
\bibinfo{author}{\bibfnamefont{L.}~\bibnamefont{{\v S}mejkal}},
  \bibinfo{author}{\bibfnamefont{R.}~\bibnamefont{Gonz{\' a}lez-Hern{\'
  a}ndez}}, \bibnamefont{and}
  \bibinfo{author}{\bibfnamefont{J.}~\bibnamefont{Sinova}},
  \bibinfo{journal}{arXiv:1901.00445}  (\bibinfo{year}{2019}).

\bibitem[{\citenamefont{Ederer and Spaldin}(2007)}]{Ederer2007}
\bibinfo{author}{\bibfnamefont{C.}~\bibnamefont{Ederer}} \bibnamefont{and}
  \bibinfo{author}{\bibfnamefont{N.~A.} \bibnamefont{Spaldin}},
  \bibinfo{journal}{Phys. Rev. B} \textbf{\bibinfo{volume}{76}},
  \bibinfo{pages}{214404} (\bibinfo{year}{2007}).

\bibitem[{\citenamefont{Spaldin et~al.}(2008)\citenamefont{Spaldin, Fiebig, and
  Mostovoy}}]{Spaldin2008}
\bibinfo{author}{\bibfnamefont{N.~A.} \bibnamefont{Spaldin}},
  \bibinfo{author}{\bibfnamefont{M.}~\bibnamefont{Fiebig}}, \bibnamefont{and}
  \bibinfo{author}{\bibfnamefont{M.}~\bibnamefont{Mostovoy}},
  \bibinfo{journal}{J. Phys.: Condensed Matter} \textbf{\bibinfo{volume}{20}},
  \bibinfo{pages}{434203} (\bibinfo{year}{2008}).

\bibitem[{\citenamefont{Spaldin et~al.}(2013)\citenamefont{Spaldin, Fechner,
  Bousquet, Balatsky, and Nordstr\"om}}]{Spaldin2013}
\bibinfo{author}{\bibfnamefont{N.~A.} \bibnamefont{Spaldin}},
  \bibinfo{author}{\bibfnamefont{M.}~\bibnamefont{Fechner}},
  \bibinfo{author}{\bibfnamefont{E.}~\bibnamefont{Bousquet}},
  \bibinfo{author}{\bibfnamefont{A.}~\bibnamefont{Balatsky}}, \bibnamefont{and}
  \bibinfo{author}{\bibfnamefont{L.}~\bibnamefont{Nordstr\"om}},
  \bibinfo{journal}{Phys. Rev. B} \textbf{\bibinfo{volume}{88}},
  \bibinfo{pages}{094429} (\bibinfo{year}{2013}).

\bibitem[{\citenamefont{Hayami et~al.}(2014)\citenamefont{Hayami, Kusunose, and
  Motome}}]{Hayami2014}
\bibinfo{author}{\bibfnamefont{S.}~\bibnamefont{Hayami}},
  \bibinfo{author}{\bibfnamefont{H.}~\bibnamefont{Kusunose}}, \bibnamefont{and}
  \bibinfo{author}{\bibfnamefont{Y.}~\bibnamefont{Motome}},
  \bibinfo{journal}{Phys. Rev. B} \textbf{\bibinfo{volume}{90}},
  \bibinfo{pages}{024432} (\bibinfo{year}{2014}).

\bibitem[{\citenamefont{Hayami et~al.}(2018)\citenamefont{Hayami, Yatsushiro,
  Yanagi, and Kusunose}}]{Hayami2018}
\bibinfo{author}{\bibfnamefont{S.}~\bibnamefont{Hayami}},
  \bibinfo{author}{\bibfnamefont{M.}~\bibnamefont{Yatsushiro}},
  \bibinfo{author}{\bibfnamefont{Y.}~\bibnamefont{Yanagi}}, \bibnamefont{and}
  \bibinfo{author}{\bibfnamefont{H.}~\bibnamefont{Kusunose}},
  \bibinfo{journal}{Phys. Rev. B} \textbf{\bibinfo{volume}{98}},
  \bibinfo{pages}{165110} (\bibinfo{year}{2018}).

\bibitem[{\citenamefont{Watanabe and Yanase}(2018)}]{Watanabe2018}
\bibinfo{author}{\bibfnamefont{H.}~\bibnamefont{Watanabe}} \bibnamefont{and}
  \bibinfo{author}{\bibfnamefont{Y.}~\bibnamefont{Yanase}},
  \bibinfo{journal}{Phys. Rev. B} \textbf{\bibinfo{volume}{98}},
  \bibinfo{pages}{245129} (\bibinfo{year}{2018}).

\bibitem[{\citenamefont{Higo et~al.}(2018)\citenamefont{Higo, Man, Gopman, Wu,
  Koretsune, van' Erve, Kabanov, Rees, Li, Suzuki et~al.}}]{Higo2018}
\bibinfo{author}{\bibfnamefont{T.}~\bibnamefont{Higo}},
  \bibinfo{author}{\bibfnamefont{H.}~\bibnamefont{Man}},
  \bibinfo{author}{\bibfnamefont{D.~B.} \bibnamefont{Gopman}},
  \bibinfo{author}{\bibfnamefont{L.}~\bibnamefont{Wu}},
  \bibinfo{author}{\bibfnamefont{T.}~\bibnamefont{Koretsune}},
  \bibinfo{author}{\bibfnamefont{O.~M.} \bibnamefont{van' Erve}},
  \bibinfo{author}{\bibfnamefont{Y.~P.} \bibnamefont{Kabanov}},
  \bibinfo{author}{\bibfnamefont{D.}~\bibnamefont{Rees}},
  \bibinfo{author}{\bibfnamefont{Y.}~\bibnamefont{Li}},
  \bibinfo{author}{\bibfnamefont{M.-T.} \bibnamefont{Suzuki}},
  \bibnamefont{et~al.}, \bibinfo{journal}{Nature photonics}
  \textbf{\bibinfo{volume}{12}}, \bibinfo{pages}{73} (\bibinfo{year}{2018}).

\bibitem[{\citenamefont{Th{\"o}le et~al.}(2016)\citenamefont{Th{\"o}le,
  Fechner, and Spaldin}}]{Thoele2016}
\bibinfo{author}{\bibfnamefont{F.}~\bibnamefont{Th{\"o}le}},
  \bibinfo{author}{\bibfnamefont{M.}~\bibnamefont{Fechner}}, \bibnamefont{and}
  \bibinfo{author}{\bibfnamefont{N.~A.} \bibnamefont{Spaldin}},
  \bibinfo{journal}{Phys. Rev. B} \textbf{\bibinfo{volume}{93}},
  \bibinfo{pages}{195167} (\bibinfo{year}{2016}).

\bibitem[{\citenamefont{Nakatsuji et~al.}(2015)\citenamefont{Nakatsuji,
  Kiyohara, and Higo}}]{Nakatsuji2015}
\bibinfo{author}{\bibfnamefont{S.}~\bibnamefont{Nakatsuji}},
  \bibinfo{author}{\bibfnamefont{N.}~\bibnamefont{Kiyohara}}, \bibnamefont{and}
  \bibinfo{author}{\bibfnamefont{T.}~\bibnamefont{Higo}},
  \bibinfo{journal}{Nature} \textbf{\bibinfo{volume}{527}},
  \bibinfo{pages}{212} (\bibinfo{year}{2015}).

\bibitem[{\citenamefont{Kiyohara et~al.}(2016)\citenamefont{Kiyohara, Tomita,
  and Nakatsuji}}]{Kiyohara2016}
\bibinfo{author}{\bibfnamefont{N.}~\bibnamefont{Kiyohara}},
  \bibinfo{author}{\bibfnamefont{T.}~\bibnamefont{Tomita}}, \bibnamefont{and}
  \bibinfo{author}{\bibfnamefont{S.}~\bibnamefont{Nakatsuji}},
  \bibinfo{journal}{Phys. Rev. Applied} \textbf{\bibinfo{volume}{5}},
  \bibinfo{pages}{064009} (\bibinfo{year}{2016}).

\bibitem[{\citenamefont{Nayak et~al.}(2016)\citenamefont{Nayak, Fischer, Sun,
  Yan, Karel, Komarek, Shekhar, Kumar, Schnelle, K{\" u}bler
  et~al.}}]{Nayak2016}
\bibinfo{author}{\bibfnamefont{A.~K.} \bibnamefont{Nayak}},
  \bibinfo{author}{\bibfnamefont{J.~E.} \bibnamefont{Fischer}},
  \bibinfo{author}{\bibfnamefont{Y.}~\bibnamefont{Sun}},
  \bibinfo{author}{\bibfnamefont{B.}~\bibnamefont{Yan}},
  \bibinfo{author}{\bibfnamefont{J.}~\bibnamefont{Karel}},
  \bibinfo{author}{\bibfnamefont{A.~C.} \bibnamefont{Komarek}},
  \bibinfo{author}{\bibfnamefont{C.}~\bibnamefont{Shekhar}},
  \bibinfo{author}{\bibfnamefont{N.}~\bibnamefont{Kumar}},
  \bibinfo{author}{\bibfnamefont{W.}~\bibnamefont{Schnelle}},
  \bibinfo{author}{\bibfnamefont{J.}~\bibnamefont{K{\" u}bler}},
  \bibnamefont{et~al.}, \bibinfo{journal}{Sci. Adv.}
  \textbf{\bibinfo{volume}{2}}, \bibinfo{pages}{e1501870}
  (\bibinfo{year}{2016}).

\bibitem[{\citenamefont{Ikhlas et~al.}(2017)\citenamefont{Ikhlas, Tomita,
  Koretsune, Suzuki, Nishio-Hamane, Arita, Otani, and Nakatsuji}}]{Ikhlas2017}
\bibinfo{author}{\bibfnamefont{M.}~\bibnamefont{Ikhlas}},
  \bibinfo{author}{\bibfnamefont{T.}~\bibnamefont{Tomita}},
  \bibinfo{author}{\bibfnamefont{T.}~\bibnamefont{Koretsune}},
  \bibinfo{author}{\bibfnamefont{M.-T.} \bibnamefont{Suzuki}},
  \bibinfo{author}{\bibfnamefont{D.}~\bibnamefont{Nishio-Hamane}},
  \bibinfo{author}{\bibfnamefont{R.}~\bibnamefont{Arita}},
  \bibinfo{author}{\bibfnamefont{Y.}~\bibnamefont{Otani}}, \bibnamefont{and}
  \bibinfo{author}{\bibfnamefont{S.}~\bibnamefont{Nakatsuji}},
  \bibinfo{journal}{Nature Phys.} \textbf{\bibinfo{volume}{13}},
  \bibinfo{pages}{1085} (\bibinfo{year}{2017}).

\bibitem[{\citenamefont{Blatt and Weisskopf}(1991)}]{Blatt1991}
\bibinfo{author}{\bibfnamefont{J.~M.} \bibnamefont{Blatt}} \bibnamefont{and}
  \bibinfo{author}{\bibfnamefont{V.~F.} \bibnamefont{Weisskopf}},
  \emph{\bibinfo{title}{Theoretical Nuclear Physics}}
  (\bibinfo{publisher}{Dover Publications, New York}, \bibinfo{year}{1991}).

\bibitem[{\citenamefont{Kusunose}(2008)}]{Kusunose2008}
\bibinfo{author}{\bibfnamefont{H.}~\bibnamefont{Kusunose}},
  \bibinfo{journal}{J. Phys. Soc. Jpn.} \textbf{\bibinfo{volume}{77}},
  \bibinfo{pages}{064710} (\bibinfo{year}{2008}).

\bibitem[{\citenamefont{Varshalovich}(1988)}]{Varshalovich1988}
\bibinfo{author}{\bibfnamefont{D.~A.} \bibnamefont{Varshalovich}},
  \emph{\bibinfo{title}{Quantum Theory Of Angular Momemtum}}
  (\bibinfo{publisher}{World Scientific Pub Co Inc}, \bibinfo{year}{1988}).

\bibitem[{\citenamefont{Th\"ole et~al.}(2016)\citenamefont{Th\"ole, Fechner,
  and Spaldin}}]{Thole2016}
\bibinfo{author}{\bibfnamefont{F.}~\bibnamefont{Th\"ole}},
  \bibinfo{author}{\bibfnamefont{M.}~\bibnamefont{Fechner}}, \bibnamefont{and}
  \bibinfo{author}{\bibfnamefont{N.~A.} \bibnamefont{Spaldin}},
  \bibinfo{journal}{Phys. Rev. B} \textbf{\bibinfo{volume}{93}},
  \bibinfo{pages}{195167} (\bibinfo{year}{2016}).

\bibitem[{\citenamefont{Hahn et~al.}(2002)\citenamefont{Hahn, Shmueli, Wilson,
  and Prince}}]{internationaltables2002}
\bibinfo{author}{\bibfnamefont{T.}~\bibnamefont{Hahn}},
  \bibinfo{author}{\bibfnamefont{U.}~\bibnamefont{Shmueli}},
  \bibinfo{author}{\bibfnamefont{A.~J.~C.} \bibnamefont{Wilson}},
  \bibnamefont{and} \bibinfo{author}{\bibfnamefont{E.}~\bibnamefont{Prince}},
  \emph{\bibinfo{title}{International tables for crystallography}}
  (\bibinfo{publisher}{KLUWER ACADEMIC PUBLISHERS}, \bibinfo{year}{2002}).

\bibitem[{\citenamefont{Bertaut}(1968)}]{Bertaut1968}
\bibinfo{author}{\bibfnamefont{E.~F.} \bibnamefont{Bertaut}},
  \bibinfo{journal}{Acta Crystallographica Section A}
  \textbf{\bibinfo{volume}{24}}, \bibinfo{pages}{217} (\bibinfo{year}{1968}).

\bibitem[{\citenamefont{Bertaut}(1981)}]{Bertaut1981}
\bibinfo{author}{\bibfnamefont{E.}~\bibnamefont{Bertaut}},
  \bibinfo{journal}{Journal of Magnetism and Magnetic Materials}
  \textbf{\bibinfo{volume}{24}}, \bibinfo{pages}{267} (\bibinfo{year}{1981}).

\bibitem[{\citenamefont{Wills et~al.}(2006)\citenamefont{Wills, Zhitomirsky,
  Canals, Sanchez, Bonville, R{\' e}otier, and Yaouanc}}]{Wills2006}
\bibinfo{author}{\bibfnamefont{A.~S.} \bibnamefont{Wills}},
  \bibinfo{author}{\bibfnamefont{M.~E.} \bibnamefont{Zhitomirsky}},
  \bibinfo{author}{\bibfnamefont{B.}~\bibnamefont{Canals}},
  \bibinfo{author}{\bibfnamefont{J.~P.} \bibnamefont{Sanchez}},
  \bibinfo{author}{\bibfnamefont{P.}~\bibnamefont{Bonville}},
  \bibinfo{author}{\bibfnamefont{P.~D.~d.} \bibnamefont{R{\' e}otier}},
  \bibnamefont{and} \bibinfo{author}{\bibfnamefont{A.}~\bibnamefont{Yaouanc}},
  \bibinfo{journal}{J. Phys.: Condens. Matter} \textbf{\bibinfo{volume}{18}},
  \bibinfo{pages}{L37} (\bibinfo{year}{2006}).

\bibitem[{\citenamefont{Machida et~al.}(2010)\citenamefont{Machida, Nakatsuji,
  Onoda, Tayama, and Sakakibara}}]{Machida2010}
\bibinfo{author}{\bibfnamefont{Y.}~\bibnamefont{Machida}},
  \bibinfo{author}{\bibfnamefont{S.}~\bibnamefont{Nakatsuji}},
  \bibinfo{author}{\bibfnamefont{S.}~\bibnamefont{Onoda}},
  \bibinfo{author}{\bibfnamefont{T.}~\bibnamefont{Tayama}}, \bibnamefont{and}
  \bibinfo{author}{\bibfnamefont{T.}~\bibnamefont{Sakakibara}},
  \bibinfo{journal}{Nature} \textbf{\bibinfo{volume}{463}},
  \bibinfo{pages}{210} (\bibinfo{year}{2010}).

\bibitem[{\citenamefont{Nakatsuji et~al.}(2011)\citenamefont{Nakatsuji,
  Machida, Ishikawa, Onoda, Karaki, Tayama, and Sakakibara}}]{Nakatsuji2011}
\bibinfo{author}{\bibfnamefont{S.}~\bibnamefont{Nakatsuji}},
  \bibinfo{author}{\bibfnamefont{Y.}~\bibnamefont{Machida}},
  \bibinfo{author}{\bibfnamefont{J.~J.} \bibnamefont{Ishikawa}},
  \bibinfo{author}{\bibfnamefont{S.}~\bibnamefont{Onoda}},
  \bibinfo{author}{\bibfnamefont{Y.}~\bibnamefont{Karaki}},
  \bibinfo{author}{\bibfnamefont{T.}~\bibnamefont{Tayama}}, \bibnamefont{and}
  \bibinfo{author}{\bibfnamefont{T.}~\bibnamefont{Sakakibara}},
  \bibinfo{journal}{Journal of Physics: Conference Series}
  \textbf{\bibinfo{volume}{320}}, \bibinfo{pages}{012056}
  (\bibinfo{year}{2011}).

\bibitem[{\citenamefont{Gaudet et~al.}(2016)\citenamefont{Gaudet, Ross,
  Kermarrec, Butch, Ehlers, Dabkowska, and Gaulin}}]{Gaudet2016}
\bibinfo{author}{\bibfnamefont{J.}~\bibnamefont{Gaudet}},
  \bibinfo{author}{\bibfnamefont{K.~A.} \bibnamefont{Ross}},
  \bibinfo{author}{\bibfnamefont{E.}~\bibnamefont{Kermarrec}},
  \bibinfo{author}{\bibfnamefont{N.~P.} \bibnamefont{Butch}},
  \bibinfo{author}{\bibfnamefont{G.}~\bibnamefont{Ehlers}},
  \bibinfo{author}{\bibfnamefont{H.~A.} \bibnamefont{Dabkowska}},
  \bibnamefont{and} \bibinfo{author}{\bibfnamefont{B.~D.}
  \bibnamefont{Gaulin}}, \bibinfo{journal}{Phys. Rev. B}
  \textbf{\bibinfo{volume}{93}}, \bibinfo{pages}{064406}
  (\bibinfo{year}{2016}).

\bibitem[{\citenamefont{Yaouanc et~al.}(2013)\citenamefont{Yaouanc, Dalmas~de
  R\'{e}otier, Bonville, Hodges, Glazkov, Keller, Sikolenko, Bartkowiak, Amato,
  Baines et~al.}}]{Yaouanc2013}
\bibinfo{author}{\bibfnamefont{A.}~\bibnamefont{Yaouanc}},
  \bibinfo{author}{\bibfnamefont{P.}~\bibnamefont{Dalmas~de R\'{e}otier}},
  \bibinfo{author}{\bibfnamefont{P.}~\bibnamefont{Bonville}},
  \bibinfo{author}{\bibfnamefont{J.~A.} \bibnamefont{Hodges}},
  \bibinfo{author}{\bibfnamefont{V.}~\bibnamefont{Glazkov}},
  \bibinfo{author}{\bibfnamefont{L.}~\bibnamefont{Keller}},
  \bibinfo{author}{\bibfnamefont{V.}~\bibnamefont{Sikolenko}},
  \bibinfo{author}{\bibfnamefont{M.}~\bibnamefont{Bartkowiak}},
  \bibinfo{author}{\bibfnamefont{A.}~\bibnamefont{Amato}},
  \bibinfo{author}{\bibfnamefont{C.}~\bibnamefont{Baines}},
  \bibnamefont{et~al.}, \bibinfo{journal}{Phys. Rev. Lett.}
  \textbf{\bibinfo{volume}{110}}, \bibinfo{pages}{127207}
  (\bibinfo{year}{2013}).

\bibitem[{\citenamefont{Pomjakushina et~al.}(2015)\citenamefont{Pomjakushina,
  Pomjakushin, Rolfs, Karpinski, and Conder}}]{Pomjakushina2015}
\bibinfo{author}{\bibfnamefont{E.}~\bibnamefont{Pomjakushina}},
  \bibinfo{author}{\bibfnamefont{V.}~\bibnamefont{Pomjakushin}},
  \bibinfo{author}{\bibfnamefont{K.}~\bibnamefont{Rolfs}},
  \bibinfo{author}{\bibfnamefont{J.}~\bibnamefont{Karpinski}},
  \bibnamefont{and} \bibinfo{author}{\bibfnamefont{K.}~\bibnamefont{Conder}},
  \bibinfo{journal}{Inorg. Chem.} \textbf{\bibinfo{volume}{54}},
  \bibinfo{pages}{9092} (\bibinfo{year}{2015}).

\bibitem[{\citenamefont{Yamaura et~al.}(2012)\citenamefont{Yamaura, Ohgushi,
  Ohsumi, Hasegawa, Yamauchi, Sugimoto, Takeshita, Tokuda, Takata, Udagawa
  et~al.}}]{Yamaura2012}
\bibinfo{author}{\bibfnamefont{J.}~\bibnamefont{Yamaura}},
  \bibinfo{author}{\bibfnamefont{K.}~\bibnamefont{Ohgushi}},
  \bibinfo{author}{\bibfnamefont{H.}~\bibnamefont{Ohsumi}},
  \bibinfo{author}{\bibfnamefont{T.}~\bibnamefont{Hasegawa}},
  \bibinfo{author}{\bibfnamefont{I.}~\bibnamefont{Yamauchi}},
  \bibinfo{author}{\bibfnamefont{K.}~\bibnamefont{Sugimoto}},
  \bibinfo{author}{\bibfnamefont{S.}~\bibnamefont{Takeshita}},
  \bibinfo{author}{\bibfnamefont{A.}~\bibnamefont{Tokuda}},
  \bibinfo{author}{\bibfnamefont{M.}~\bibnamefont{Takata}},
  \bibinfo{author}{\bibfnamefont{M.}~\bibnamefont{Udagawa}},
  \bibnamefont{et~al.}, \bibinfo{journal}{Phys. Rev. Lett.}
  \textbf{\bibinfo{volume}{108}}, \bibinfo{pages}{247205}
  (\bibinfo{year}{2012}).

\bibitem[{\citenamefont{Poole et~al.}(2007)\citenamefont{Poole, Wills, and
  Leli\'{e}vre-Berna}}]{Poole2007}
\bibinfo{author}{\bibfnamefont{A.}~\bibnamefont{Poole}},
  \bibinfo{author}{\bibfnamefont{A.~S.} \bibnamefont{Wills}}, \bibnamefont{and}
  \bibinfo{author}{\bibfnamefont{E.}~\bibnamefont{Leli\'{e}vre-Berna}},
  \bibinfo{journal}{J. Phys.: Condens. Matter} \textbf{\bibinfo{volume}{19}},
  \bibinfo{pages}{452201} (\bibinfo{year}{2007}).

\bibitem[{\citenamefont{Taira et~al.}(2003)\citenamefont{Taira, Wakeshima,
  Hinatsu, Tobo, and Ohoyama}}]{Taira2003}
\bibinfo{author}{\bibfnamefont{N.}~\bibnamefont{Taira}},
  \bibinfo{author}{\bibfnamefont{M.}~\bibnamefont{Wakeshima}},
  \bibinfo{author}{\bibfnamefont{Y.}~\bibnamefont{Hinatsu}},
  \bibinfo{author}{\bibfnamefont{A.}~\bibnamefont{Tobo}}, \bibnamefont{and}
  \bibinfo{author}{\bibfnamefont{K.}~\bibnamefont{Ohoyama}},
  \bibinfo{journal}{J. Solid Chem.} \textbf{\bibinfo{volume}{176}},
  \bibinfo{pages}{165} (\bibinfo{year}{2003}).

\bibitem[{\citenamefont{Wiebe et~al.}(2004)\citenamefont{Wiebe, Gardner, Kim,
  Luke, Wills, Gaulin, Greedan, Swainson, Qiu, and Jones}}]{Wiebe2004}
\bibinfo{author}{\bibfnamefont{C.~R.} \bibnamefont{Wiebe}},
  \bibinfo{author}{\bibfnamefont{J.~S.} \bibnamefont{Gardner}},
  \bibinfo{author}{\bibfnamefont{S.-J.} \bibnamefont{Kim}},
  \bibinfo{author}{\bibfnamefont{G.~M.} \bibnamefont{Luke}},
  \bibinfo{author}{\bibfnamefont{A.~S.} \bibnamefont{Wills}},
  \bibinfo{author}{\bibfnamefont{B.~D.} \bibnamefont{Gaulin}},
  \bibinfo{author}{\bibfnamefont{J.~E.} \bibnamefont{Greedan}},
  \bibinfo{author}{\bibfnamefont{I.}~\bibnamefont{Swainson}},
  \bibinfo{author}{\bibfnamefont{Y.}~\bibnamefont{Qiu}}, \bibnamefont{and}
  \bibinfo{author}{\bibfnamefont{C.~Y.} \bibnamefont{Jones}},
  \bibinfo{journal}{Phys. Rev. Lett.} \textbf{\bibinfo{volume}{93}},
  \bibinfo{pages}{076403} (\bibinfo{year}{2004}).

\bibitem[{\citenamefont{Mirebeau et~al.}(2005)\citenamefont{Mirebeau, Apetrei,
  Rodr\'{i}guez-Carvajal, Bonville, Forget, Colson, Glazkov, Sanchez, Isnard,
  and Suard}}]{Mirebeau2005}
\bibinfo{author}{\bibfnamefont{I.}~\bibnamefont{Mirebeau}},
  \bibinfo{author}{\bibfnamefont{A.}~\bibnamefont{Apetrei}},
  \bibinfo{author}{\bibfnamefont{J.}~\bibnamefont{Rodr\'{i}guez-Carvajal}},
  \bibinfo{author}{\bibfnamefont{P.}~\bibnamefont{Bonville}},
  \bibinfo{author}{\bibfnamefont{A.}~\bibnamefont{Forget}},
  \bibinfo{author}{\bibfnamefont{D.}~\bibnamefont{Colson}},
  \bibinfo{author}{\bibfnamefont{V.}~\bibnamefont{Glazkov}},
  \bibinfo{author}{\bibfnamefont{J.~P.} \bibnamefont{Sanchez}},
  \bibinfo{author}{\bibfnamefont{O.}~\bibnamefont{Isnard}}, \bibnamefont{and}
  \bibinfo{author}{\bibfnamefont{E.}~\bibnamefont{Suard}},
  \bibinfo{journal}{Phys. Rev. Lett.} \textbf{\bibinfo{volume}{94}},
  \bibinfo{pages}{246402} (\bibinfo{year}{2005}).

\bibitem[{\citenamefont{Sazonov et~al.}(2013)\citenamefont{Sazonov, Gukasov,
  Cao, Bonville, Ressouche, Decorse, and Mirebeau}}]{Sazonov2013}
\bibinfo{author}{\bibfnamefont{A.~P.} \bibnamefont{Sazonov}},
  \bibinfo{author}{\bibfnamefont{A.}~\bibnamefont{Gukasov}},
  \bibinfo{author}{\bibfnamefont{H.~B.} \bibnamefont{Cao}},
  \bibinfo{author}{\bibfnamefont{P.}~\bibnamefont{Bonville}},
  \bibinfo{author}{\bibfnamefont{E.}~\bibnamefont{Ressouche}},
  \bibinfo{author}{\bibfnamefont{C.}~\bibnamefont{Decorse}}, \bibnamefont{and}
  \bibinfo{author}{\bibfnamefont{I.}~\bibnamefont{Mirebeau}},
  \bibinfo{journal}{Phys. Rev. B} \textbf{\bibinfo{volume}{88}},
  \bibinfo{pages}{184428} (\bibinfo{year}{2013}).

\bibitem[{\citenamefont{Disseler et~al.}(2015)\citenamefont{Disseler, Borchers,
  Brooks, Mundy, Moyer, Hillsberry, Thies, Tenne, Heron, Holtz
  et~al.}}]{Disseler2015}
\bibinfo{author}{\bibfnamefont{S.~M.} \bibnamefont{Disseler}},
  \bibinfo{author}{\bibfnamefont{J.~A.} \bibnamefont{Borchers}},
  \bibinfo{author}{\bibfnamefont{C.~M.} \bibnamefont{Brooks}},
  \bibinfo{author}{\bibfnamefont{J.~A.} \bibnamefont{Mundy}},
  \bibinfo{author}{\bibfnamefont{J.~A.} \bibnamefont{Moyer}},
  \bibinfo{author}{\bibfnamefont{D.~A.} \bibnamefont{Hillsberry}},
  \bibinfo{author}{\bibfnamefont{E.~L.} \bibnamefont{Thies}},
  \bibinfo{author}{\bibfnamefont{D.~A.} \bibnamefont{Tenne}},
  \bibinfo{author}{\bibfnamefont{J.}~\bibnamefont{Heron}},
  \bibinfo{author}{\bibfnamefont{M.~E.} \bibnamefont{Holtz}},
  \bibnamefont{et~al.}, \bibinfo{journal}{Phys. Rev. Lett.}
  \textbf{\bibinfo{volume}{114}}, \bibinfo{pages}{217602}
  (\bibinfo{year}{2015}).

\bibitem[{\citenamefont{Mu\~{n}oz et~al.}(2000)\citenamefont{Mu\~{n}oz, Alonso,
  Mart\'{i}nez-Lope, Cas\'{i}s, Mart\'{i}nez, and
  Fern\'{a}ndez-D\'{i}az}}]{Munoz2000}
\bibinfo{author}{\bibfnamefont{A.}~\bibnamefont{Mu\~{n}oz}},
  \bibinfo{author}{\bibfnamefont{J.~A.} \bibnamefont{Alonso}},
  \bibinfo{author}{\bibfnamefont{M.~J.} \bibnamefont{Mart\'{i}nez-Lope}},
  \bibinfo{author}{\bibfnamefont{M.~T.} \bibnamefont{Cas\'{i}s}},
  \bibinfo{author}{\bibfnamefont{J.~L.} \bibnamefont{Mart\'{i}nez}},
  \bibnamefont{and} \bibinfo{author}{\bibfnamefont{M.~T.}
  \bibnamefont{Fern\'{a}ndez-D\'{i}az}}, \bibinfo{journal}{Phys. Rev. B}
  \textbf{\bibinfo{volume}{62}}, \bibinfo{pages}{9498} (\bibinfo{year}{2000}).

\bibitem[{\citenamefont{Brown and Chatterji}(2006)}]{Brown2006}
\bibinfo{author}{\bibfnamefont{P.~J.} \bibnamefont{Brown}} \bibnamefont{and}
  \bibinfo{author}{\bibfnamefont{T.}~\bibnamefont{Chatterji}},
  \bibinfo{journal}{J. Phys.: Condens. Matter} \textbf{\bibinfo{volume}{18}},
  \bibinfo{pages}{10085} (\bibinfo{year}{2006}).

\bibitem[{\citenamefont{Fabr\'{e}ges et~al.}(2008)\citenamefont{Fabr\'{e}ges,
  Mirebeau, Bonville, Petit, Lebras-Jasmin, Forget, Andr\`{e}, and
  Pailh\`{e}s}}]{Fabreges2008}
\bibinfo{author}{\bibfnamefont{X.}~\bibnamefont{Fabr\'{e}ges}},
  \bibinfo{author}{\bibfnamefont{I.}~\bibnamefont{Mirebeau}},
  \bibinfo{author}{\bibfnamefont{P.}~\bibnamefont{Bonville}},
  \bibinfo{author}{\bibfnamefont{S.}~\bibnamefont{Petit}},
  \bibinfo{author}{\bibfnamefont{G.}~\bibnamefont{Lebras-Jasmin}},
  \bibinfo{author}{\bibfnamefont{A.}~\bibnamefont{Forget}},
  \bibinfo{author}{\bibfnamefont{G.}~\bibnamefont{Andr\`{e}}},
  \bibnamefont{and}
  \bibinfo{author}{\bibfnamefont{S.}~\bibnamefont{Pailh\`{e}s}},
  \bibinfo{journal}{Phys. Rev. B} \textbf{\bibinfo{volume}{78}},
  \bibinfo{pages}{214422} (\bibinfo{year}{2008}).

\end{thebibliography}

\end{document}